\documentclass[final,3p]{elsarticle}

 \usepackage{graphics}

 \usepackage{graphicx}

\usepackage{caption}
\usepackage{float}
\usepackage{subcaption}

\usepackage{amssymb}
\usepackage{amsmath}
\usepackage{color}
\usepackage{soul}          
\usepackage{ulem}
\usepackage{tikz}
\usepackage{makecell}
\usepackage{algorithm}
\usepackage{algpseudocode}
\usetikzlibrary{shapes.geometric, arrows}

\tikzstyle{normal} = [rectangle, rounded corners, minimum width=3cm, minimum height=1cm,text centered, draw=black, fill=red!30,text width=10em]

\tikzstyle{robust} = [trapezium, trapezium left angle=70, trapezium right angle=110, minimum width=3cm, minimum height=1cm, text centered, draw=black, fill=blue!30,text width=4em]

\usepackage[T1]{fontenc}
\usepackage{hyperref}
\hypersetup{
colorlinks=true
}
\usepackage{epstopdf}
\usepackage{xcolor}

\biboptions{sort&compress}

\begin{document}

\begin{frontmatter}

\title{
{{Assessment} of background noise properties in time and time-frequency domains in the context of
vibration-based local damage detection in real environment}}

\author[label1]{Katarzyna Skowronek}
  \author[label3]{Tomasz Barszcz} 
    \author[label4]{Jerome Antoni}  \author[label2]{Rados{\l}aw Zimroz} 
\author[label1]{Agnieszka Wy\l oma\'nska}

\address[label1]{Faculty of Pure and Applied Mathematics, Hugo Steinhaus Center, Wroclaw University of Science and Technology, Wyspianskiego 27, 50-370 Wroclaw, Poland}

\address[label3]{Faculty of Mechanical Engineering and Robotics, AGH University of Science and Technology, Al. Mickiewicza 30, 30-059 Krakow, Poland}
\address[label4]{Univ Lyon, INSA Lyon, LVA, EA677, 69621 Villeurbanne, France}

\address[label2]{Faculty of Geoengineering, Mining and Geology, Wroclaw University of Science and Technology, Na Grobli 15, 50-421 Wroclaw, Poland }

\begin{abstract}
{Any measurement in condition monitoring applications is associated with disturbing noise. Till now, most of the diagnostic procedures have assumed the Gaussian distribution for the noise. This paper shares a novel perspective to the problem of local damage detection. The acquired vector of observations is considered as an additive mixture of signal of interest (SOI) and noise with strongly non-Gaussian, heavy-tailed properties, that masks the SOI. The distribution properties of  the background noise influence the selection of tools used for the signal analysis, particularly for local damage detection. Thus, it is extremely important to recognize and identify possible non-Gaussian behavior of the noise. The problem considered here is more general than the classical goodness-of-fit testing. {The} paper highlights the important  role  of variance,  as most of the methods for signal analysis are based on the assumption of the finite-variance distribution of the underlying signal. The finite variance  assumption is crucial but implicit to most indicators  used in condition monitoring, (such as the root-mean-square value, the power spectral density, the kurtosis, the spectral correlation, etc.), in view that infinite variance implies moments higher than $2$ are also infinite. The problem is demonstrated based on three  popular types of non-Gaussian distributions observed for real vibration signals. We demonstrate how the properties of noise distribution in the time domain may change by its  transformations to the time-frequency domain (spectrogram).  Additionally, we propose a procedure to check the presence of the infinite-variance of the background noise. Our investigations are illustrated using simulation studies and real vibration signals from various machines.}

\end{abstract}

 
 \date{}

\end{frontmatter}

\section{Introduction}
In many technical systems, the measurement of any physical variable is performed to acquire some important information about object or process. Let us call it signal of interest (SOI). Although measurement systems are very advanced, the informative components may be noisy because of the presence of other stronger sources, non-informative in a given context. 
 Thus, a natural concept of signal analysis is related to pre-processing (de-noising) to improve the signal-to-noise ratio (SNR) and to better detect the SOI.  Signal de-noising can be done directly in the time domain (TD), in the frequency domain (FD) or time-frequency domain (TFD).  It is intuitive, that before removing the noise from the signal, one needs to identify the noise component properties as many techniques are developed under the specific  assumptions  of the corresponding distribution  
 \cite{Lin20049,Zhao2017129,Yu2022}. 
 One of the most common assumptions for de-noising techniques is  the Gaussian distribution of the background noise \cite{Yu2022}.  However, in everyday practice, various techniques for de-noising are used without checking if this assumption is fulfilled. On the other side, this may bring significant consequences because  applying methods dedicated for specific cases (i.e. assuming Gaussian distribution) may return inaccurate results for signals not satisfying the assumed properties. Some researchers indicated this problem and discuss the limitations of classical  methods for signals not satisfying the assumed properties, see \cite{zak20161932,Wodecki201989,Wodecki2021,LIU2017311,PANCALDI2023109975}.
 
 The other issue that needs to be highlighted is related to the properties of the signal distribution  after its transformation to other domains, like TFD. Also, some of the de-noising techniques act on the signals in other domains (time domain, spectral, wavelet coefficients, bi-frequency map) \cite{Schmidt2021,Lin20049,Wodecki201989,Obuchowski2014389,Wodecki2021}, so such issue is important also in this context. Even if the signal in TD has required properties,  after the transformation to other domains they may change drastically. The perfect example is the transformation of Gaussian distributed signal to TFD (spectrogram), see \cite{nadine1}. After this transformation, we obtain the signal with generalized $\chi^2$ distribution, which has different properties from the Gaussian one. Thus, the de-noising techniques dedicated to the signals with Gaussian properties may not give the expected results when applying them to time-frequency representation.  In the literature, one may find some interesting research where this issue is considered and the distribution properties change (after signal transformations to other domains) are discussed \cite{nadine1,mik,aw1,aw2,aw3}. 

Let us move one step forward and consider the problem of periodic/cyclic behavior identification of the signal. Here, we assume that the signal is a mixture of SOI and the background noise. This is a typical model  used for local damage detection in rotating elements such as bearings or gears. The target is to detect the SOI hidden in background noise. In the case of finite-variance distributed  noise (e.g. Gaussian), the analysis of random signals can be done by using classical auto-dependence measures. The most common example is the auto-covariance (ACVF) or auto-correlation function (ACF) and the classical approaches for periodic/cyclic behavior identification utilize such functions. There are techniques where signals are analyzed in TD as well as TFD \cite{gardner2006cyclostationarity,hurd2007periodically,randall2001relationship}, see also \cite{csc2,csc1} for new approaches.

However, when applying such measures, one needs to take into account they are properly defined only for finite-variance distributed signals. Applying their sample versions  to the signals from infinite-variance distribution is inappropriate, and the obtained results may not give expected information. This problem was discussed in our previous research \cite{kruczek2020detect,kruczek2021generalized,nowicki2020local,zak2019periodically} but also other authors analyze this issue and propose dedicated techniques for impulsive signals \cite{cyclo4,mssp1,cyclic_corre4,cyclic_corre5,cyclic_corre6} and highlight the small efficiency of the classical methods, see e.g. \cite{BARSZCZ2011431,BARSZCZ20091352,URBANEK20121782,URBANEK201713}. There were also proposed  transformations that can help to make the non-Gaussian signals closer to Gaussian, see for instance \cite{borghesani2017cs2} where the authors showed that a simple logarithmic transform on the squared envelope had an excellent stabilizing effect before computing its Fourier transform. Preliminary knowledge about the noise properties can help to avoid inappropriate conclusions resulting from the use of wrong tools for signal analysis and may help to select more adequate techniques, like robust estimators of auto-dependence measures dedicated for impulsive signals \cite{pz1,pz2,pz3,pz4,pz5,kend3}  or auto-dependence measures defined  for signals  with some specific non-Gaussian distributions \cite{floc2,floc3,nowicka1,ded2,wylomanska2015codifference}.
 We note, similar as for the de-noising techniques, some authors test the classical methods for local damage detection also for signals with impulsive behavior \cite{zak2017measures,Schmidt2021,Wodecki2021}. The analysis presented in the mentioned above bibliography positions clearly indicates the limitations of the classical techniques for extreme cases.
 
 In view of the above discussion, we note that the preliminary analysis of the background noise properties is extremely important for  selection of appropriate tools for signal analysis which, in turn, is necessary to obtain reliable results. As in this paper we discuss the problem in the context of periodic/cyclic behavior identification for signal-based local damage detection, our attention is paid to the identification if the distribution of the signal has finite or infinite variance. {More precisely, we assess the probabilistic properties (in the mean of finite or infinite variance) of the background noise that affects the properties of the signal (being a mixture of the SOI and the noise) itself.} We note, the considered problem is much more general than the classical goodness-of-fit testing \cite{test1,test2,test3}, i.e. testing if the underlying signal has a given theoretical distribution. It is worth noting that the distribution identification is the last step of the analysis and may have less importance than the preliminary knowledge of the distribution category (here in the context of finite and infinite variance).  
 
 In the literature, one can find interesting approaches when some specific properties of given data are tested \cite{infvartest,heavytest}.  In our  previous research we also analyzed the problem of heavy-tailed behavior recognition \cite{krzysiek0,krzysiek1,Iskander} but it was considered for specific classes of distributions. In this paper, we present the broader perspective and discuss the problem in the context of any non-Gaussian distributions with possible infinite variance. 
 
In this paper, we recall three most popular non-Gaussian distributions with possible infinite-variance (depending on the parameters). We explain the selection of these distributions in the context of local damage detection in rotating machines. We discuss the problem of finite- and infinite-variance distribution  of the  signals in TD and TFD. More precisely, we highlight that finite-variance property  of the corresponding distribution in TD may not be transferred to TFD (here spectrogram), that may have the significant importance for further analysis. We propose a visual test for checking  if the variance is finite. The variance is a key parameter for classical auto-dependence measures applications.  As we are working with real vibration signals with complex spectral content, all analyses are performed in TFD. Thus, the assessment of the probabilistic properties of a random noise component is done for some wider frequency range, not just an arbitrary selected sub-signal for a given frequency band taken from the spectrogram. In order to achieve this, we propose an objective, automatic procedure based on the mentioned visual test. To illustrate the problem and results of our investigations, we present several exemplary real vibration measurements from different machines and demonstrate their probabilistic properties in  TFD (spectrogram). We also provide deep simulation study to highlight the importance of the research topic presented in this paper. 

The rest of the paper is organized as follows.
In Section \ref{pr_formulation} we formulate the considered problem indicating  two perspectives, practical problem of local damage detection and probabilistic point of view. Next, in Section \ref{sec3} we recall three considered non-Gaussian distributions considered here as the general classes with possible finite and infinite variances. Then, we discuss the distribution of the signals  transferred to TFD (spectrogram). Here, we indicate four separate categories of distributions that are crucial for selection of the appropriate tools for signals analysis in TD and TFD. In Section \ref{sec3} we also propose an automatic procedure for the infinite variance behavior analysis. In Section \ref{simul} we analyze the simulated signals from three considered distributions and demonstrate their probabilistic properties  in TD and TFD. Moreover, for the simulated signals, we demonstrate the procedure for infinite variance testing. In Section \ref{real} we analyze four real signals from different machines and demonstrate their differences in the context of the probabilistic properties by using the proposed methodology. The last section concludes the paper.

\section{Problem formulation}\label{pr_formulation}
{Let us assume that acquired signal consists of two main components: informative signal (SOI) and non-informative component (called simple noise). If SNR is high, the presence of noise may be neglected. However, in real applications,  especially in local fault detection problems, the SOI may be completely hidden in the noise. As the SOI (in our case) has two specific properties (impulsiveness and periodicity), there are plenty of techniques that allow its detection, even under strong domination of the noise. However, in most of the cases the crucial assumption needs to be fulfilled, namely the noise should be Gaussian distributed. Unfortunately, in various applications, the noise exhibits non-Gaussian impulsive behavior.  What does it exactly mean? Each noise that is not  Gaussian distributed simply may be considered as non-Gaussian. However, in local damage detection, a very important properties of the SOI is its impulsive character and many techniques are based on this property. If we consider non-Gaussian heavy-tailed distributed noise, the situation becomes very complicated. Heavy-tailed distribution means that in the realization of the process some very large (positive and negative) values (i.e. outliers) may appear. An example of such a heavy-tailed noise is an impulsive process. According to our research, sources of such impulsive behavior may be related to specific processes performed by machine (cutting, crushing, milling, drilling, compression, etc) \cite{kuba10,nowicki2020local, borghesani2017cs2,wodecki2019impulsive,Lahrache201721,Yu2013155}, may be related to external completely random disturbances, disturbances during data transmission or even numerical problems during data processing \cite{Mauricio2020}. In other words, the problem of impulsive noise  may appear in practice in many situations.

However, the mixture of the SOI and impulsive (non-Gaussian infinite-variance) noise excludes {impulsive criteria that involve moments equal or greater than $2$ (such as kurtosis, smoothness index,  etc.)} \cite{HebdaSobkowicz2020,hebda2022infogram,Wodecki2021,kruczek2020detect,antoni2006spectral}) as SOI detectors. {It is worth highlighting that infinite variance of a given distribution  implies the higher moments are also infinite. Thus, for instance, the infinite-variance distributed signal has also infinite kurtosis (which is based on the second and third moment).} The other possibility to detect SOI  is to measure the auto-dependence of the signal. However, as it was mentioned,  in some cases the classical auto-dependence measures (ACF, ACVF) cannot be used. {According to Wiener-Khinchine theorem, the infinite second moment also does not allow using of the power spectral density (PSD), which is one of the most versatile tools in condition monitoring.} Thus, the important issue is to identify the probabilistic properties of the noise and confirm that the classical methods can be applied. But how to check the nature of the noise component? In real applications, the spectral content of the acquired signal may be complicated. Some sources may be deterministic, so the signal should be decomposed first to obtain only the random part. Thus, we suggest not testing properties of the signal in TD, but in TFD and to use spectrogram for signal decomposition. 
It is already known that if signal in TD has a Gaussian distribution, in TFD (spectrogram) it has a generalized $\chi^2$ distribution \cite{nadine1}. The situation is much complicated when the nose is non-Gaussian heavy-tailed distributed. In that case, after transformation to TFD, the distribution is unknown. Moreover, even if the level of non-Gaussianity in TD is negligible (signal is almost Gaussian), after transformation to TFD, properties of the signal can be very different. It may happen, that even if we could apply the classical methods (like ACVF or ACF) in TD, we should not use them for signal in time-frequency representation.  

In order to deal with this problem, one needs appropriate tools to investigate properties of the noise and constrains regarding usage of classical methods. We propose to use a visual test to check if the variance of the corresponding distribution is finite and, therefore, if classical methods can be used. As we are working in TFD, we analyze so called sub-signals - narrowband signals for each frequency bin. Decision-making based on arbitrary selected sub-signal may be significantly biased. Thus, we adopted the visual test for finite variance  and proposed an automatic procedure applied to wide frequency band. As test based on empirical cumulative fourth moment results in the plot of statistics that should converge to constant value for finite-variance distributed signals. The analysis of simulated signals from non-Gaussian distributions confirms the efficiency of the proposed algorithm for the considered issue. }

\section{Theoretical background}\label{sec3}
As it was mentioned, the background noise of the real vibration signals very often exhibits non-Gaussian  heavy-tailed behavior. There are plenty of distributions belonging to this class, however in  this paper we consider three general exemplary cases. 

\subsection{Non-Gaussian distributions of random signals}
The first considered distribution is the $\alpha$-stable one $ \mathcal{S}(\alpha,\sigma)$. This distribution has been successfully used in condition monitoring applications by various authors, see e.g. \cite{Yu2013155,zak3,zak56,8726382,7819828,6911199,6324624}.  In this paper, we consider the symmetric version of the $\alpha-$ stable distribution defined through the characteristic function \cite{stable1,stable2}
\begin{eqnarray}
  \Phi_X(x)=\mathbf{E}\exp\left\{iXx\right\}=\exp{\left( -\sigma^{\alpha} |x|^{\alpha}\right)}, \quad \quad x \in \mathbb{R}
\end{eqnarray}
 where $0<\alpha\leq2$ is the stability index and $\sigma>0$ is the scale parameter. For $\alpha=2$ the $\alpha-$stable distribution reduces to the Gaussian one, and thus it can be considered as a generalization of this classical distribution. The symmetric $\alpha-$stable distribution has no closed-form probability density function (PDF) and cumulative distribution function (CDF). The only exception is the Gaussian distribution (that is, for $\alpha=2$) and the Cauchy distribution (that is, for $\alpha=1$). The stability index is responsible for the heaviness of this distribution's tail, $1-F_X(x)\sim x^{-\alpha}$ ($F_X(\cdot)$ is a CDF of random variable $X$), i.e. the smaller $\alpha$ the probability of large values is  much higher. For $\alpha<2$, the variance  of $\alpha-$stable distribution is infinite.

As the second example of non-Gaussian heavy-tailed distribution, we consider the symmetric Pareto one. We selected this distribution because, similar as $\alpha-$stable one,  it can have finite- and infinite-variance and exhibits the power-law behavior. On the other side, in contrast to $\alpha-$stable distribution, its PDF and CDF are given in explicit forms which makes statistical inference simpler in this case.  The PDF of the symmetric Pareto distribution $\mathcal{P}(\gamma,\lambda)$ is given by \cite{wylomanska2016impulsive} 
 \begin{equation}
        f_X(x)= \frac{\gamma \lambda^\gamma}{2(|x|+ \lambda)^{\gamma+1}}, \quad \quad x \in \mathbb{R}.
\end{equation}
Similarly, as for the stability index in the $\alpha-$stable case, the $\gamma>0$ parameter is responsible for the heaviness of this distribution's tail.  The $\lambda>0$ parameter is responsible for the scale of a random variable $X$. When $\gamma>2$, the double Pareto distribution has finite variance. In that case $\gamma\leq 2$, the variance is infinite. 

The last considered  distribution is the t location-scale  $\mathcal{T}(\nu,\delta)$ defined through the PDF \cite{student1,student2}
    
    \begin{equation}
    f_X(x) = \frac{\Gamma\left( \frac{\nu+1}{2}\right)}{\delta \sqrt{\nu \pi} \Gamma\left( \frac{\nu}{2}\right)} \left[ \frac{\nu + \frac{x^2}{\delta^2}}{\nu}\right]^{-\left(  \frac{\nu+1}{2}\right)},\quad\quad x\in \mathbb{R},
    \end{equation}
where $\Gamma(\cdot)$ is the gamma function, $\nu >0$ is a  shape parameter and $\delta >0$ is a scale parameter. The variance for t-location scale distribution is only defined for $\nu >2$. Otherwise, it is infinite.   When $\nu$ tends to infinity, then t-location scale distribution tends to a Gaussian distribution. 

\begin{figure}[H]
    \centering
    
    \includegraphics[width=.7\textwidth]{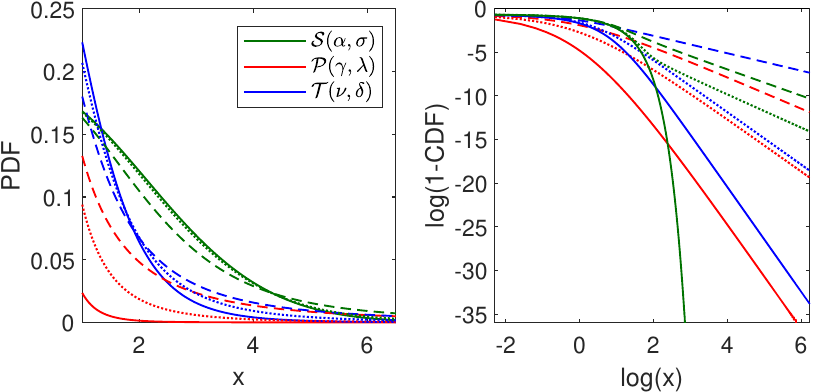}
    \caption{The PDFs (for positive arguments) and the corresponding distributions' tails (in log-log scales) for $\alpha-$stable (green lines), symmetric Pareto (red lines) and t location-scale (blue lines) distributions. The solid lines correspond to the finite-variance cases (i.e. for $\alpha=2$, $\gamma=\nu=6$). The dashed lines correspond to the infinite-variance cases (i.e. for $\alpha=\gamma=\nu=1.5$). The dotted lines correspond to the intermediate cases (i.e. $\alpha=1.9$, $\gamma=\nu=3$). 
    }
    \label{fig2}
\end{figure}

To demonstrate differences between the analyzed distributions, in Fig. \ref{fig2} we present their probability density functions  (for positive arguments) and the corresponding distributions' tails (in log-log scales).  The green lines correspond to the $\alpha-$stable distribution, the red lines to the symmetric Pareto distribution, while the blue lines to the t location-scale distribution. In each case, the solid lines correspond to the finite-variance cases (i.e., for $\alpha=2$, $\gamma=\nu=6$) while the dashed lines correspond to the infinite-variance cases (i.e., for $\alpha=\gamma=\nu=1.5$). The intermediate cases, i.e. $\alpha=1.9$, $\gamma=\nu=3$ are marked in dotted lines. In the right panel one can see that the tails of non-Gaussian distributions are "heavier" than in the Gaussian case (i.e. for $\alpha-$stable with $\alpha=2$) where the large observations can occur with smaller probability than for other cases. Moreover, for the finite-variance cases (solid lines) the distributions' tails are "lighter' than for infinite-variances (dashed lines and dotted green line). 

\subsection{Distribution of non-Gaussian signals in time-frequency representation}
Since most methods for local damage detection operate in time-frequency representation, in this part we discuss the distribution of the signals transformed into time-frequency map. It is recalled that the signal is a mixture of the SOI  -- assumed nonstationary when analyzed in the time-frequency domain -- and of background noise (that is reasonably assumed stationary), and the latter only is the subject of analysis of the present paper. . As for the time-frequency representation,  we propose to use the  spectrogram (the square of the absolute value of short time Fourier transform, STFT) in view of that many methods in condition monitoring are rooted on this representation, {e.g. the spectrogram and its reassigned versions such as the synchrosqueezing transform, the Welch's estimator of the power spectral density and of the spectral correlation, etc.

We recall for the signal $x_1,x_2,\cdots,x_n$ the STFT is defined as follows
 \begin{equation}
 \label{STFT1}
        STFT(t,f) = ST_R(t,f) + i ST_I(t,f) = \sum_{m=1}^{n} x_{m} w(t - m)e^{-i2\pi f \frac{m}{n}},
\end{equation}
\noindent where $ST_R(t,f)$ and $ST_I(t,f)$ are real and imaginary parts of STFT respectively,  $w(\cdot)$ is window, $t \in T$ is time point and $f \in \mathcal{F}$ is frequency. Real and imaginary parts of STFT can be expressed as

 \begin{equation}
 \label{STR}
        ST_R(t,f) = \sum_{m=1}^{n} x_{m} w(t - m )\text{cos}(-2\pi f\frac{m}{n}),~~
        ST_I(t,f) =  \sum_{m=1}^{n} x_{m} w(t - m)\text{sin}(-2\pi f  \frac{m}{n}),
\end{equation}

\noindent where $ST_R(t,f)$ and $ST_I(t,f)$ are real and imaginary parts of STFT, respectively.
The spectrogram $S(\cdot,\cdot)$ is a square of the absolute value of STFT

 \begin{equation}
 \label{STFT}
        S(t,f) = |STFT(t,f)|^2 = ST_R(t,f)^2 +ST_I(t,f)^2.
\end{equation}

As it was highlighted in \cite{nadine1} that if the signal $x_1,x_2,\cdots,x_n$ represents  samples of independent observations from the  zero-mean Gaussian distribution, then for any $f\in \mathcal{F}$ the vectors $ST_R(\cdot,f)$ and $ST_I(\cdot,f)$ comes also from a Gaussian distribution with mean equal to zero. The exact formulas for the variances of the Gaussian distributions corresponding  to the samples given in (\ref{STR})  and their theoretical covariance are  presented in \cite{nadine1}. 

According to \cite{nadine1}, under the assumption the signal $x_1,x_2,\cdots,x_n$ of independent observations comes from the centered Gaussian distribution, the $S(\cdot,f)$ for any $f\in \mathcal{F}$ is a sample from the so-called  generalized $\chi^2$ distribution defined thought the PDF, \cite{MatPro92}
\begin{align}
\label{chi2pdf}
    p_{\theta,\beta}(x) = \frac{1}{(2\beta)^\frac{\theta}{2} \Gamma\left(\frac{\theta}{2}\right)} x^{\frac{\theta-2}{2}}\text{exp} \left( - \frac{x}{2\beta}\right), ~x>0,
\end{align}
where parameter $\theta\in \mathbb{N}$ is called a number of degrees of freedom, $\beta>0$ is the scale parameter and $\Gamma(\cdot)$ is a gamma function. The generalized $\chi^2$ distribution has also strong connection with gamma distribution, see e.g., \cite{grebenkov,MatPro92}. One can show that  under the assumption of Gaussian distributed signal, the spectrogram given in Eq. (\ref{STFT}), is a quadratic form of Gaussian variables. Based on the theory of Gaussian quadratic forms \cite{MatPro92}, in \cite{nadine1} the authors calculated the parameters of generalized $\chi^2$ distribution of $S(\cdot,f)$ for any $f\in \mathcal{F}$. {The parameters $\theta$ and $\beta$ are expressed in means of variances of $ST_R(\cdot,f)$ and $ST_I(\cdot,f)$ and their covariance}. If we denote the corresponding theoretical variances as $\sigma_R^2$ and $\sigma_I^2$, respectively and  covariance as  $\sigma_{ST}$, then  the theoretical values of the parameters of $\chi^2$ distribution  corresponding to $S(\cdot,f)$ (for any $f\in \mathcal{F}$) are
given by
\begin{eqnarray}
    \theta = \frac{(\sigma_{R}+\sigma_{I})^2}{\sigma_{R}^2+\sigma_{I}^2+2\sigma_{ST}^2}, ~   \beta = \frac{\sigma_{R}+\sigma_{I}}{2}.
\end{eqnarray}

The situation is much more complicated if the signal  $x_1,x_2,\cdots,x_n$ is not Gaussian distributed. In the case of symmetric double Pareto or t-location scale distribution, there is no analytical formula describing the distribution of the noise in the real and imaginary part of the STFT or the spectrogram. We only highlight that depending on the values of parameters responsible for the heavy-tailed behavior (i.e. $\gamma$ and $\nu$ for symmetric Pareto and t location-scale distributions, respectively), we can obtain finite- or infinite-variance distributed samples in time-frequency representations of the signal (spectrogram).

For the $\alpha$-stable distributed signals, $ST_R(\cdot,f)$ and $ST_I(\cdot,f)$ are also $\alpha$-stable distributed with the same stability index $\alpha$ as the signal $x_1,x_2,\cdots,x_n$. This follows from the probabilistic properties of the $\alpha-$stable random variables \cite{stable} and the generalized central limit theorem \cite{gnedenko}.  Using the results presented in \cite{kluppelberg1993spectral,klup1994}, where the distribution of the squared Fourier transform (periodogram) for $\alpha$-stable linear processes is discussed, we may conclude the distribution of the series $S(\cdot,f)$ for any $f\in \mathcal{F}$  belongs to the domain of attraction of $\alpha/2-$stable distribution if the random signal $x_1,x_2,\cdots,x_n$ is $\alpha-$stable distributed (with stability index $\alpha$). 

In this paper, we highlight the important  role of variance  of corresponding theoretical distribution for selection of appropriate tools for signal analysis. The information about variance existence (i.e. if the variance is finite of infinite for corresponding distribution) is crucial for further steps. Thus, in Fig. \ref{schematic} we demonstrate possible types of distributions (in the categories of finite and infinite-variance case) of signals in TD and TFD. We highlight that the finiteness of the variance of the signal distribution in TD does not guarantee its finiteness after its transformation to TFD (spectrogram).  We remind, in the class denoted as (4) there is included  the $\alpha-$stable distribution with $\alpha<2$. In the schema, we do not highlight this distribution as a separate category. In Table \ref{tab1} we present the considered distributions and ranges of their parameters (responsible for non-Gaussian behavior, i.e. $\alpha,\gamma$ and $\nu$ for $\alpha-$stable, symmetric Pareto and t location-scale distributions, respectively) corresponding to cases (1)-(4) of the schema presented in Fig. \ref{schematic}. We note, the parameters $\sigma,\lambda$ and $\delta$ for $\alpha-$stable, symmetric Pareto and t location-scale distributions, respectively, do not have influence on the non-Gaussian behavior. Thus, we do not include them in the table. Table \ref{tab1} may be useful for real signal analysis when identification of the distribution type for the background noise  may  be important for the selection of appropriate tools for local damage detection.

\begin{figure}
    \centering
    \includegraphics[width=1\textwidth]{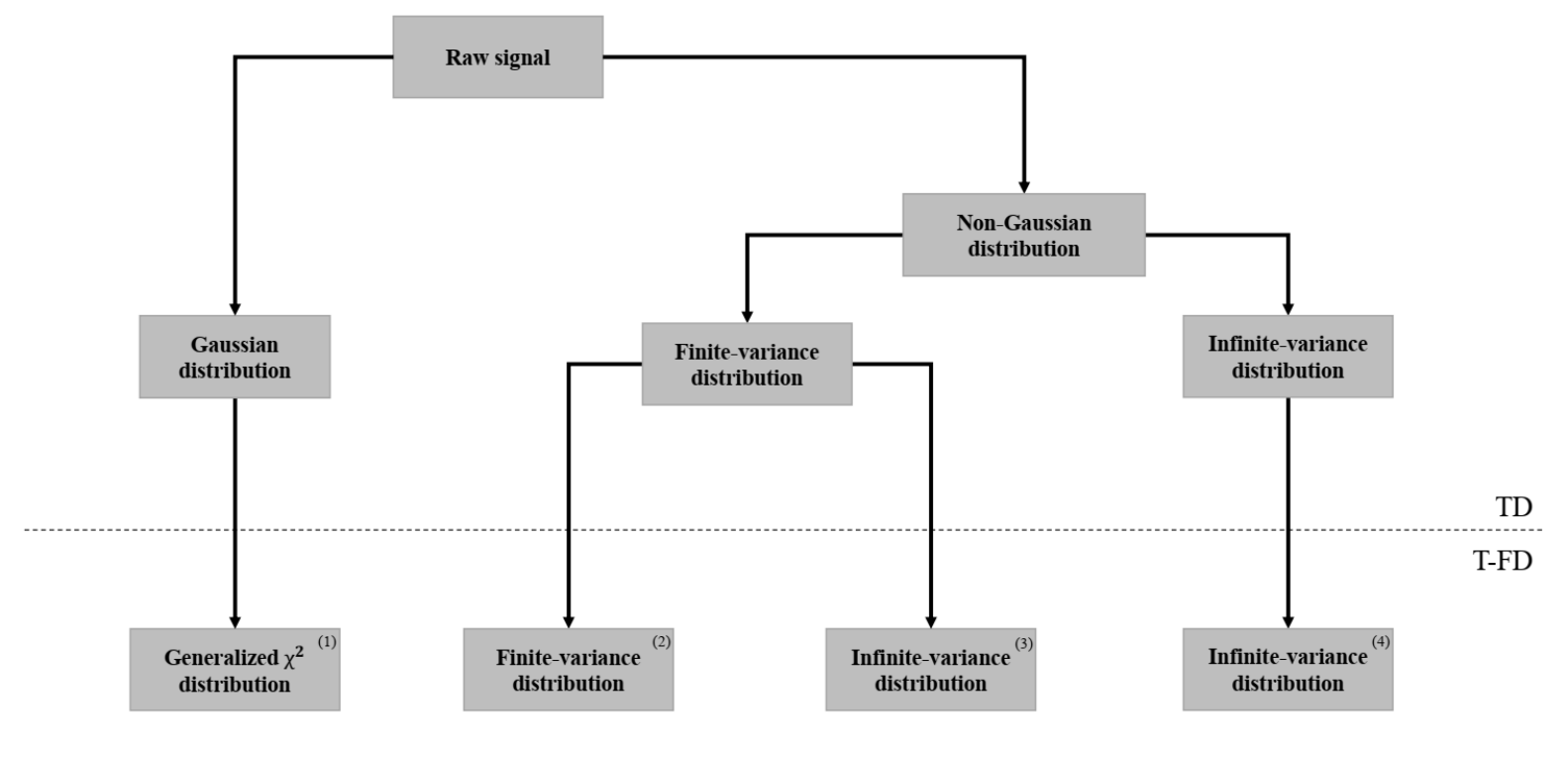}
    \caption{Possible types of distributions (in the categories of finite and infinite-variance distribution) of signals in TD and TFD.}
    \label{schematic}
\end{figure}

\begin{table}[h!]
\centering
\begin{tabular}{|l|l|l|l|l|l|}
\hline
Distribution  & Parameters & Variance & Distribution & Variance & Category  \\ 
TD & Parameters &TD& TFD& TFD & (see Fig. \ref{schematic})  \\
\hline
$\mathcal{S}(\alpha,\sigma)$  & $\alpha=2$&finite  & Generalized $\chi^2$ & finite &(1) \\ \hline$\mathcal{P}(\gamma,\lambda)$, $\mathcal{T}(\nu,\delta)$ & $\gamma> 4,\nu> 4$ &finite & unknown & finite & (2)            \\ \hline $\mathcal{P}(\gamma,\lambda)$, $\mathcal{T}(\nu,\delta)$ & $2<\gamma\leq 4, 2<\nu\leq 4$ &finite & unknown & infinite &(3)  \\ \hline
$\mathcal{P}(\gamma,\lambda)$, $\mathcal{T}(\nu,\delta)$ & $0<\gamma\leq 2,0<\nu\leq2$ & infinite&unknown & infinite  &  (4)\\ \hline
$\mathcal{S}(\alpha,\sigma)$     & $0<\alpha<2$ &infinite & domain of attraction & infinite  &(4)    \\ 
& &&of $\alpha/2-$stable &&\\\hline

\end{tabular}
\caption{The considered distributions and ranges of their parameters corresponding to the categories (1)-(4) presented in Fig. \ref{schematic} of the signals' distribution in  TD and TFD.}\label{tab1}
\end{table}

\subsection{Assessment of probabilistic properties of random signals in time and time-frequency domains}    \label{ident_time}
As it was noted, the problem considered here is much more general than the classical goodness-of-fit testing if real signal can be modeled by a given theoretical distribution. In this paper, we test if the signal belongs to the finite- or infinite-variance class of distributions without specification of the distribution, as in many cases this identification may be very difficult or even impossible. To the assessment of the infinite-variance behavior we propose to use the simple statistic, called empirical cumulative fourth moment (ECFM) that was analyzed in our previous research in the similar context. The selection of this statistic for the discussed problem is related to the fact that ECFM has a simple form, and  it exhibits completely different behavior for finite- and infinite-variance distributed signals that is a crucial point in the testing procedure. 

In our previous research \cite{krzysiek0,krzysiek1} we have discussed  the problem of discrimination between Gaussian and near-Gaussian distributions for which the variance may be infinite. The  perfect example was the $\alpha-$stable distribution with the stability index close to $2$. However, this methodology may also be applied to other distributions considered in this paper, and it can be extended  for the assessment of the probabilistic properties of the signal also in TFD. In \cite{krzysiek0} to distinguish the Gaussian and infinite-variance distribution of given data, the authors proposed to use the  ECFM statistic 
\begin{eqnarray}\label{ecfm}
C(k)=\frac{1}{k}\sum_{i=1}^k(x_i-\overline{x})^4,\quad \quad k=1,2,\cdots,n,
\end{eqnarray}
where $x_1,x_2,\cdots,x_n$ is the considered signal of independent identically distributed (i.i.d.) observations and $\overline{x}$ the corresponding sample mean.

In \cite{krzysiek0,krzysiek1} it was highlighted that  the statistic given in Eq. (\ref{ecfm}) converges to a
constant for the Gaussian distribution (or any other distribution with finite fourth moment). In practice, given a finite sample, one observes that ECFM exhibits irregular chaotic behaviour only for distributions with infinite fourth moment. In this paper,  the ECFM statistic is applied to confirm or reject the finite-variance distribution of the signal in TFD. 

In the following part of this section, we show how to parameterize the chaotic behavior of the ECFM statistic for infinite-variance distributed signals. As it was mentioned, the methods for local damage detection are mostly based on the analysis of the signals in TFD. Thus, in this paper, the procedure presented below  is applied to the time-frequency representation (spectrogram) of a given signal. Precisely, we measure the chaotic behavior of ECFM statistic for  vectors $S(\cdot,f)$ for all $f\in \mathcal{F}$ and analyse the distribution of this measure along the frequencies.  However, this algorithm can be also applied for identification of infinite-variance  behavior for signals in any other domains. 

The procedure consists of the following steps:
\begin{enumerate}
    \item The signal $x_1,x_2,\cdots,x_n$ first is transformed to TFD. In this paper, we use the spectrogram defined in Eq. (\ref{STFT}).
    \item For each $f\in \mathcal{F}$ we normalize the vector $S(\cdot,f)$ by subtracting its sample mean and by dividing by the sample conditional standard deviation on the quantile levels $0.1$ and $0.9$ (see \cite{pitera} for more details). The normalisation was performed in order to standardise the output and reduce the influence of the mean and scale parameters. Note that we followed empirical-based standardisation due to the heavy-tail nature of the data linked e.g. to infinite variance.
    \item For each vector $S(\cdot,f)$ we calculate the ECFM statistic according to Eq. (\ref{ecfm}).
    \item For each $f\in \mathcal{F}$  we identify the segments of the ECFM statistic between the jumps. In order to identify the segments, first we calculate the  increments of the ECFM statistic and then  identify their peaks, considering them as the points separating the  segments. 
    \item For each $f\in \mathcal{F}$ we select the segments of the ECFM statistic that are long enough. In our analysis, we selected the segments of minimum $10\%$ of the length of $S(\cdot,f)$. For further analysis, we take the last long segment.  The corresponding vector we denote as $D(\cdot,f)$. %We note, for each $f$ the length of this segment can have different length, we call it $d_f$.
    \item For each $f\in \mathcal{F}$  we fit the straight line to the vector $D(\cdot,f)$ using the least squares method. The estimated value of the slope parameter for frequency $f$ we denote as $a_f$.
    \item We analyze the distribution of the estimated slopes along the frequencies. If the signal in TFD has finite-variance distribution, we expect the ECFM statistic calculated for vectors $S(\cdot,f)$ stabilizes. Thus, in this case, the $a_f$ parameters are close to zero. More precisely, the distribution of $a_f$ is concentrated around zero. We expect here that the median of the slopes is close to zero and the interquartile range, IQR, is  small. 
    On the other side, if the distribution of the signal in TFD has infinite variance, then the ECFM statistic for each sub-signal $S(\cdot,f)$ exhibits chaotic behavior and  estimated $a_f$ parameters are non-zero. We expect here the median of the estimated slopes (in absolute values) is significantly higher than zero and their IQR is high. 
    \end{enumerate}
{The pseudo code of the described above procedure in presented in \ref{ap_a}, see Algorithm 1.}

After the verification if the corresponding theoretical distribution is finite- or infinite- variance, the next step is the identification of the theoretical distribution corresponding to the signal. This point may be crucial (especially for testing the Gaussian or generalized $\chi^2$ distributions) as some of the local damage detection methods are dedicated only for special distributions of given signal.  To fit the proper distribution (or to select which one is the best choice from the tested ones), we propose to use the simple visual test based on the comparison of the empirical tail and the theoretical one corresponding to the tested distribution (with the estimated parameters from considered signal). The empirical tail is defined similarly as the theoretical one, however the theoretical CDF is replaced by the empirical one, i.e. $1-\hat{F}_n(x)$. We recall,  the empirical CDF for i.i.d. signal $x_1,x_2,\cdots,x_n$ is defined as follows  \cite{normal2} 
\begin{eqnarray}\label{ecdf}
\hat{F}_{n}(x)=\frac{1}{n}\sum_{j=1}^n1\{x_j\leq x\}, 
\end{eqnarray}
where $1\{A\}$ is the indicator of a set $A$.

By comparing theoretical and empirical tails, one can conclude which distribution from tested ones is more proper for the considered signal. To confirm that the tested distribution is the best choice, we propose to use the Kolmogorov-Smirnov (KS) statistic defined as 
\begin{eqnarray}\label{kstest}
KS=\sup_{x}\left|{F}_{X}(x)-\hat{F}_n(x)\right|,
\end{eqnarray}
where $F_X(\cdot)$ is the CDF of the tested theoretical distribution with the parameters estimated from the signal and $\hat{F}_n(\cdot)$ is the empirical CDF given in Eq. (\ref{ecdf}). Small value of KS statistic indicates that the empirical distribution is close to the tested one. The KS statistic can also be used for the given distribution testing. Based on Monte Carlo simulations, we can calculate the corresponding $p_{value}$.

\section{Simulated signals analysis}\label{simul}

In this section we analyze the simulated signals, and  we present how to assess the finite and infinite-variance distribution for the i.i.d. observations in time and time-frequency domains. The methodology is demonstrated for three distributions described in Section \ref{sec3}.

In Fig. \ref{fig1} we demonstrate the exemplary simulated signals from  $\mathcal{S}(\alpha,\sigma)$, $\mathcal{P}(\gamma,\lambda)$ and $\mathcal{T}(\nu,\delta)$.  For each of the considered distribution we consider three cases of parameters responsible for heavy-tailed behavior, namely for $\alpha-$stable case we take $\alpha=2$ (Gaussian distribution), $\alpha=1.9$ and $\alpha=1.5$; for symmetric Pareto distribution we take $\gamma=6$, $\gamma=3$ and $\gamma=1.5$; for t location-scale we take $\nu=6$, $\nu=3$ and $\nu=1.5$.  The other parameters are assumed to be one. Let us emphasize that $\alpha=2$ and $\gamma=\nu=6$ in $\alpha-$stable, symmetric Pareto and t location-scale distributions, respectively, correspond to the finite-variance cases while for $\alpha=\gamma=\nu=1.5$ we have infinite-variance distributions. In the middle row of Fig. \ref{fig1} we present the intermediate cases. More precisely, for $\gamma=\nu=3$ the symmetric Pareto and t location-scale distributions have finite variances while $\alpha=1.9$ for $\alpha-$stable distribution corresponds to the infinite-variance case. These cases will be discussed in depth. 

In Fig. \ref{fig1} one can see the significant differences between finite- and infinite-variance cases. For the signals presented in the bottom row and middle-left panel, the large observations are clearly visible.

In Fig. \ref{fig3} we demonstrate the ECFM statistic calculated for the samples demonstrated in Fig. \ref{fig1}. One can see the clear differences between top and bottom rows. For the finite-variance cases (top row), the ECFM statistics tend to some constants. In the bottom row, the ECFM statistics exhibit chaotic behavior which clearly confirms the infinite-variance distributions.  The middle row—left column indicates that the considered sample is infinite-variance distributed, while the middle and right columns demonstrate the finite-variance behavior of the sample. However, it is not as clear as in the top row for the corresponding distributions.

\begin{figure}[H]
    \centering
    \begin{minipage}[t]{0.45\textwidth}
        \centering
        \includegraphics[width=0.95\textwidth]{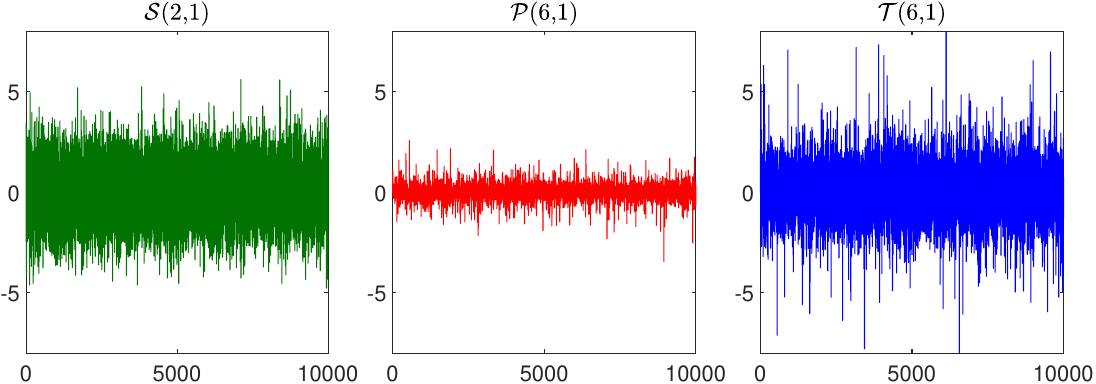} \\ 
        \includegraphics[width=0.95\textwidth]{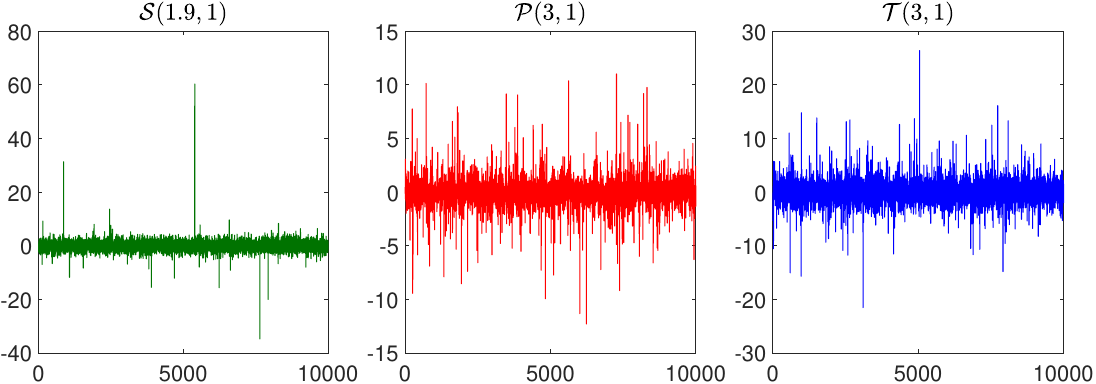} \\
        \includegraphics[width=0.95\textwidth]{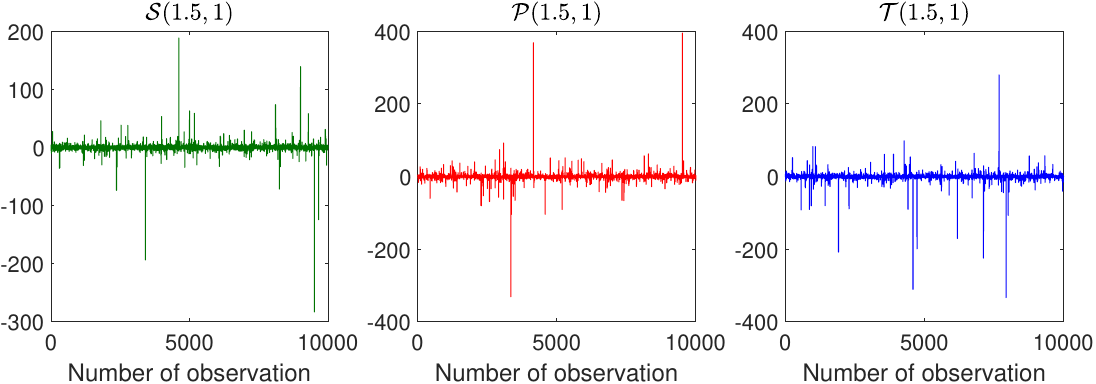}
        \caption{Examples of simulated signals from  $\mathcal{S}(\alpha,\sigma)$, $\mathcal{P}(\gamma,\lambda)$ and $\mathcal{T}(\nu,\delta)$ distributions. The green color (left columns) corresponds to the $\alpha-$stable distribution with $\alpha=2$ (top panel), $\alpha=1.9$  (middle panel) and with $\alpha=1.5$ (bottom panel). The red color (middle column) corresponds to the symmetric Pareto distribution with $\gamma=6$ (top panel), $\nu=3$ (middle panel)  and $\gamma=1.5$ (bottom panel). The blue color (right columns) corresponds to the t location-scale distribution with $\nu=6$ (top panel), $\nu=3$ (middle panel) and $\nu=1.5$ (bottom panel).   The other parameters are assumed to be one.}
    \label{fig1}
    \end{minipage}
    % \hfill
    \begin{minipage}[t]{0.45\textwidth}
        \centering
        \includegraphics[width=0.95\textwidth]{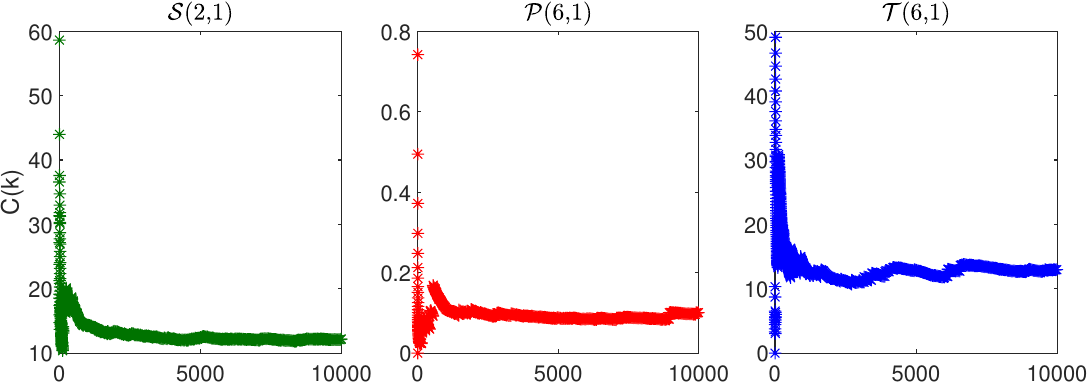} \\ 
        \includegraphics[width=0.95\textwidth]{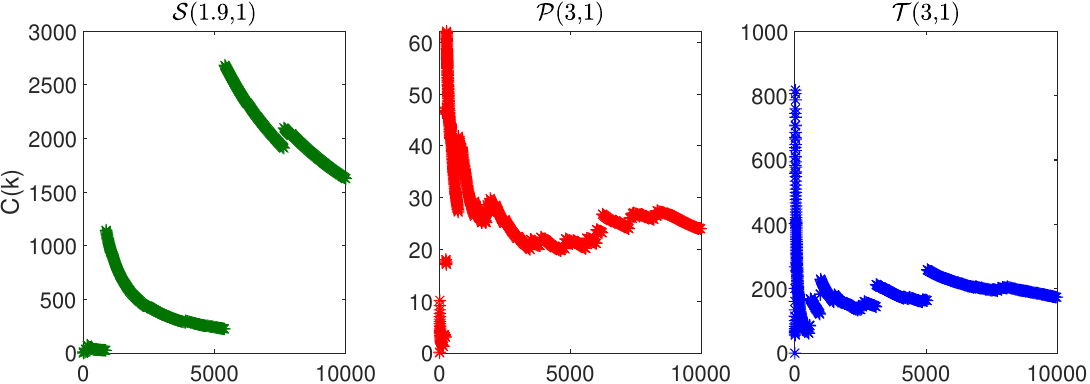} \\
        \includegraphics[width=0.95\textwidth]{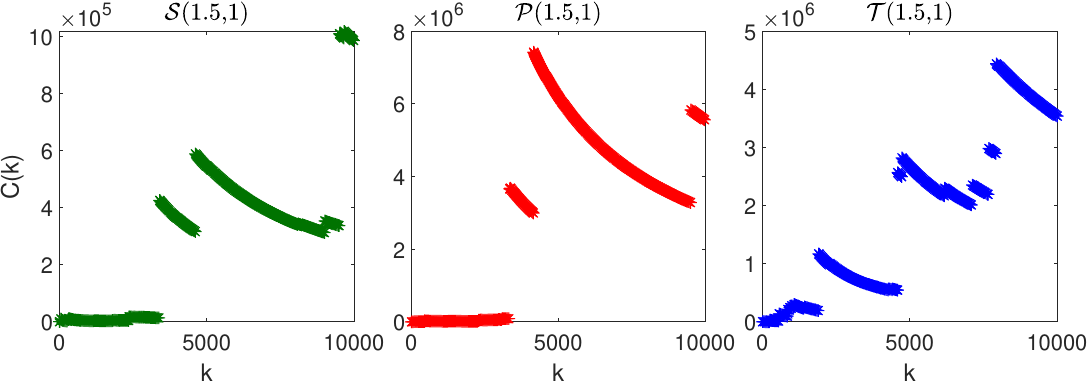}
        \caption{The ECFM statistic calculated for the signals demonstrated in Fig. \ref{fig1}. The finite-variance cases are presented in the top row, while the infinite-variance cases are demonstrated in the bottom row. In the middle row, we demonstrate the intermediate cases.}
    \label{fig3}
    \end{minipage}
\end{figure}

The spectrograms of the signals presented in Fig. \ref{fig1} are presented in Fig. \ref{fig7}.  
To demonstrate the ECFM-based methodology, for further analysis, we select a specific frequency $f=1.0963kHz$ and examine the corresponding sub-signals from the spectrogram. Because the signals in TD are represented by independent observations and do not contain any additional components, in this case the selection of $f$ is not a critical issue. However, when we analyze the real signals, the selection of the appropriate frequency bin corresponding to the noise component is a crucial point.  In Fig. \ref{fig_sub_signals} we demonstrate the sub-signals from the spectrograms presented in Fig. \ref{fig7} corresponding to the selected frequency $f$. In the middle and bottom rows, one can notice the occurrence of large values in the corresponding sub-signals. The impulsive behavior is not visible in the top row.

\begin{figure}
    \centering
    
    \includegraphics[width=.7\textwidth]{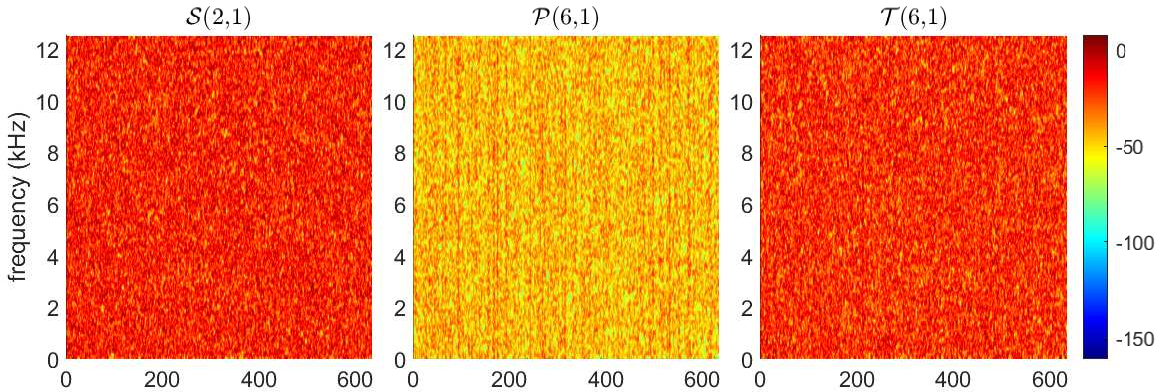}\\ \includegraphics[width=.7\textwidth]{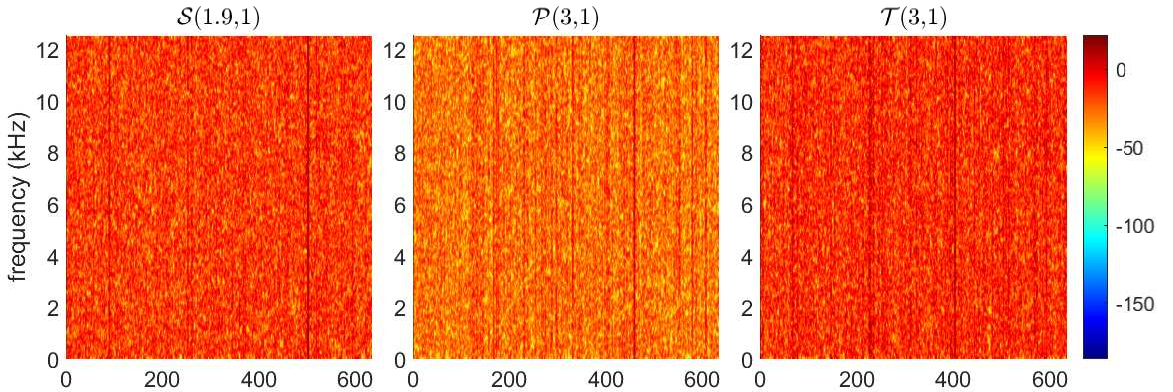}\\ \includegraphics[width=.7\textwidth]{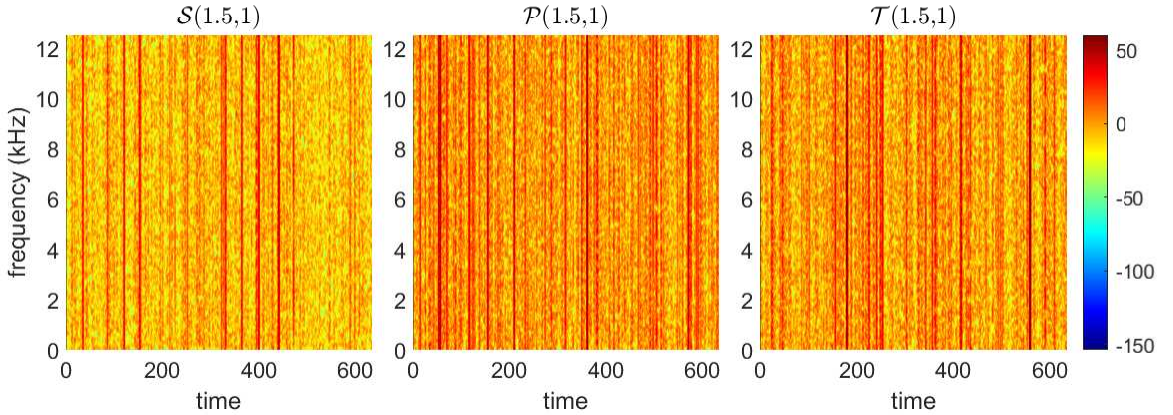}
    \caption{The spectrograms (in log scales) for simulated signals from  $\mathcal{S}(\alpha,\sigma)$, $\mathcal{P}(\gamma,\lambda)$ and $\mathcal{T}(\nu,\delta)$ distributions presented in Fig. \ref{fig1}. The left columns correspond to the $\alpha-$stable distribution with $\alpha=2$ (top panel), $\alpha=1.9$  (middle panel) and with $\alpha=1.5$ (bottom panel). The middle column corresponds to the symmetric Pareto distribution with $\gamma=6$ (top panel), $\nu=3$ (middle panel)  and $\gamma=1.5$ (bottom panel). The right columns correspond to the t location-scale distribution with $\nu=6$ (top panel), $\nu=3$ (middle panel) and $\nu=1.5$ (bottom panel).   The other parameters are assumed to be one.    }
    \label{fig7}
\end{figure}

\begin{figure}[H]
    \centering
    \begin{minipage}[t]{0.45\textwidth}
        \centering
        \centering
   
    \includegraphics[width=.95\textwidth]{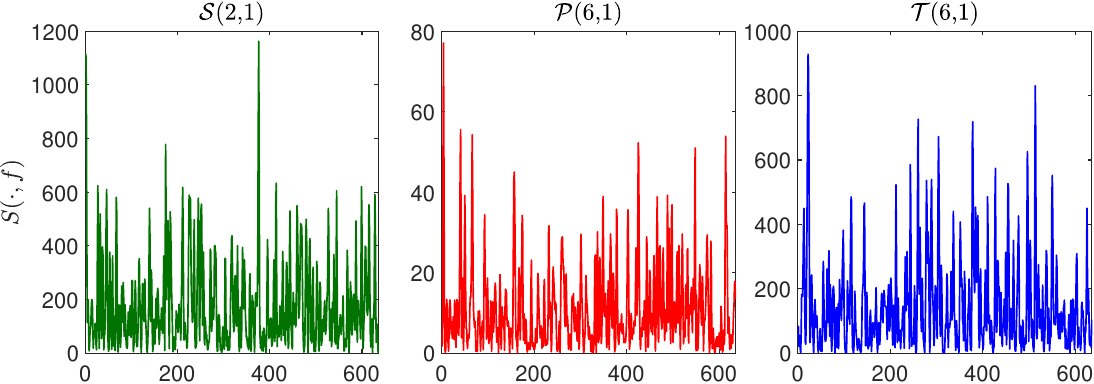}\\
    \includegraphics[width=.95\textwidth]{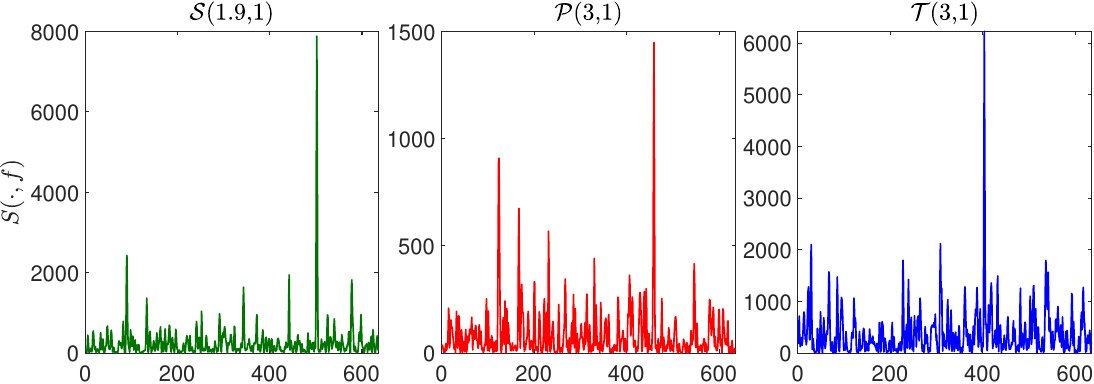}\\
    \includegraphics[width=.95\textwidth]{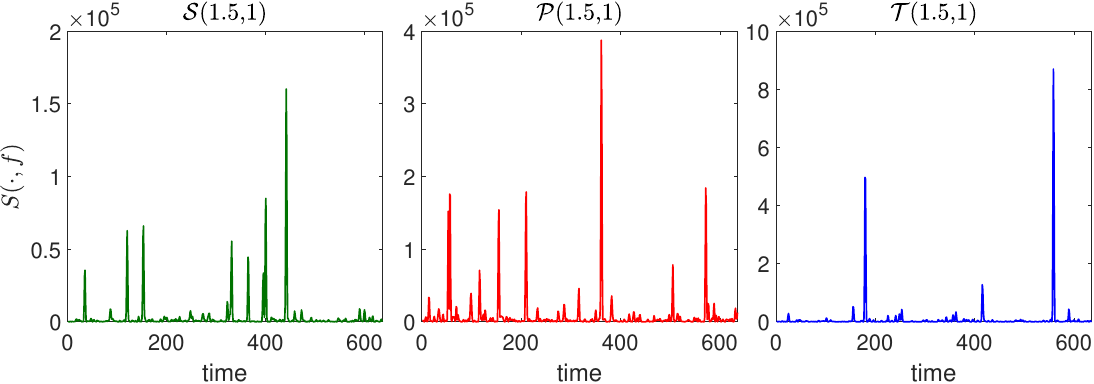}
    \caption{The sub-signals (corresponding to  $f=1.0963kHz$) from spectrograms presented in Fig. \ref{fig7} for  exemplary simulated signals from  $\mathcal{S}(\alpha,\sigma)$, $\mathcal{P}(\gamma,\lambda)$ and $\mathcal{T}(\nu,\delta)$ distributions. The green color (left columns) corresponds to the $\alpha-$stable distribution with $\alpha=2$ (top panel), $\alpha=1.9$  (middle panel) and with $\alpha=1.5$ (bottom panel). The red color (middle column) corresponds to the symmetric Pareto distribution with $\gamma=6$ (top panel), $\nu=3$ (middle panel)  and $\gamma=1.5$ (bottom panel). The blue color (right columns) corresponds to the t location-scale distribution with $\nu=6$ (top panel), $\nu=3$ (middle panel) and $\nu=1.5$ (bottom panel). The other parameters are assumed to be one.
    }
    \label{fig_sub_signals}
    \end{minipage}
    % \hfill
    \begin{minipage}[t]{0.45\textwidth}
        \centering
   
    \includegraphics[width=.95\textwidth]{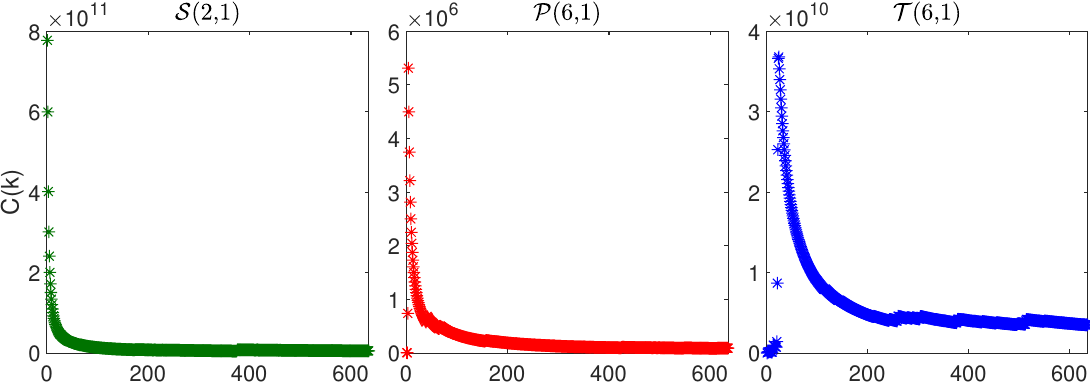}\\
    \includegraphics[width=.95\textwidth]{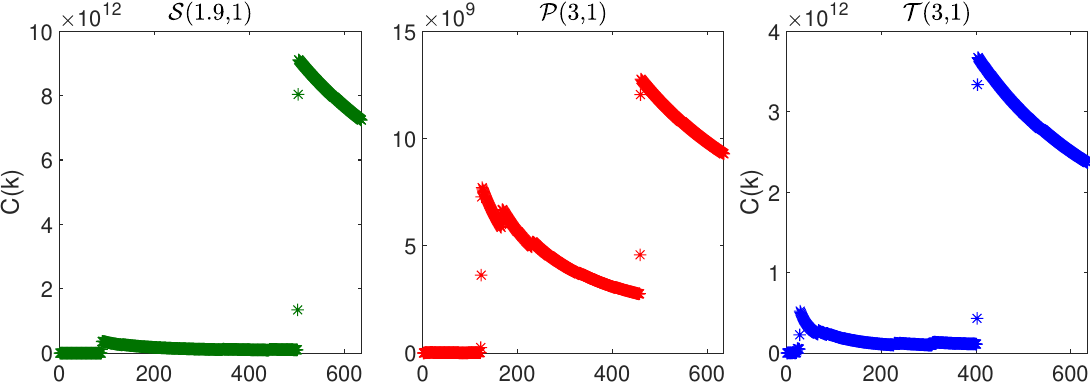}\\
    \includegraphics[width=.95\textwidth]{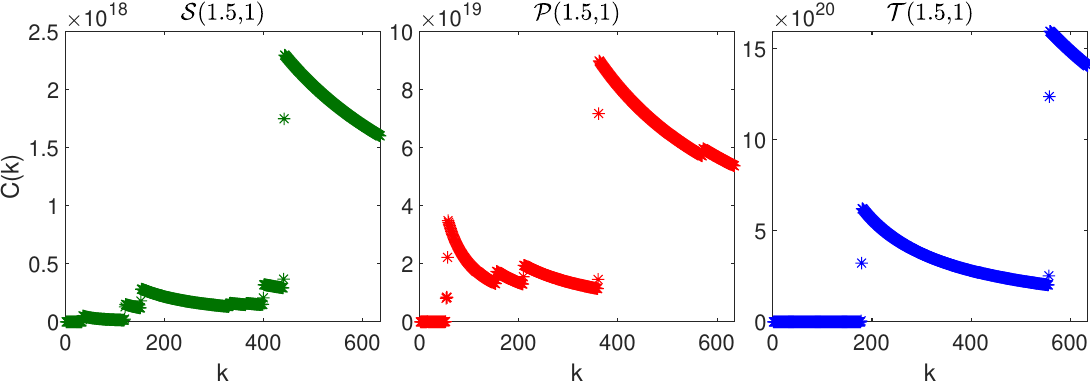}
    \caption{The ECFM statistic calculated for the sub-signals demonstrated in Fig. \ref{fig_sub_signals}. The finite-variance cases (in TFD) are presented in the top row, while the infinite-variance cases (in TFD) are demonstrated in the middle and bottom row.}
    \label{fig_ecfm_sub_sig}
    \end{minipage}
\end{figure}

The ECFM statistics for the sub-signals presented in Fig. \ref{fig_sub_signals} are demonstrated in Fig. \ref{fig_ecfm_sub_sig}. One can see in the top row that the ECFM stabilizes, which clearly indicates the finite-variance cases. In the middle row for all considered cases, the ECFM exhibits chaotic behavior, which indicates the infinite-variance distributions. It should be noted that the middle row corresponds to the $\alpha-$ stable distribution with 
$\alpha=1.9$ (left panel), the symmetric Pareto distribution with $\gamma=3$ (middle panel), and the location-scale distribution t with $\gamma=3$ (right panel). For the symmetric Pareto and the t location-scale distributions, the signals in TD were classified as finite-variance distributed; see the middle row (middle and right panels) in Fig. \ref{fig3}. This is the case, when the characteristics of the signal in TD are not transferred to TFD. In the bottom panels of Fig. \ref{fig_ecfm_sub_sig} we present the ECFM statistic for the signals classified in TD as infinite-variance distributed. The similar property we observe for the signals in TFD.

In the next part, we present the application of the described in the previous section procedure for testing of finite- and infinite variance for simulated signals from three considered distributions.
According to the algorithm, first the signals are transformed to TFD (spectrograms). Then, the ECFM is calculated for each sub-signal (see Fig. \ref{fig_ecfm_sub_sig} for exemplary sub-signals). In the next step for each sub-signal (after normalisation) the increments of ECFM statistic are calculated (see examples in Fig. \ref{incr_ecfm}) to identify the segments of the statistic between jums (see point 4. of the procedure). Finally, for each sub-signal the last long segment of ECFM  is selected, and the straight line is fitted (see point 6 of the procedure).  The selected segments of ECFM statistic and the fitted lines (marked by black lines) with estimated slopes $a_f$ are presented in Fig. \ref{segment} for sub-signals demonstrated in Fig. \ref{fig_sub_signals}. For sub-signals corresponding to the finite-variance distributions in TD (see top panels in Fig. \ref{segment}) the $\hat{a}_f$ values are close to zero. For the infinite-variance and intermediate cases (middle and bottom panels of Fig. \ref{segment}) the estimated values (in absolute values) are  higher than zero, however for the signals from $\mathcal{S}(1.5,1)$, $\mathcal{P}(1.5,1)$ and $\mathcal{T}(1.5,1)$ distributions the obtained values are significantly higher than for the intermediate cases (i.e. for $\mathcal{S}(1.9,1)$, $\mathcal{P}(3,1)$ and $\mathcal{T}(3,1)$ distributions). 

\begin{figure}[H]
    \centering
    \begin{minipage}[t]{0.45\textwidth}
        \centering
        \includegraphics[width=0.95\textwidth]{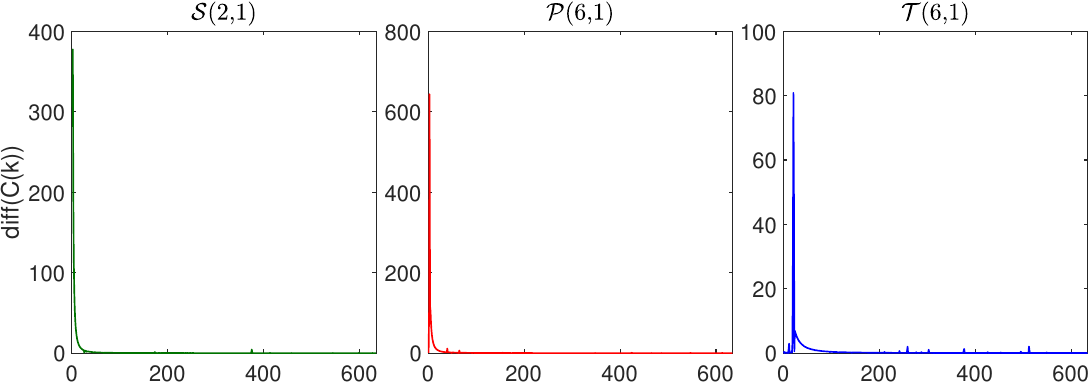} \\ 
        \includegraphics[width=0.95\textwidth]{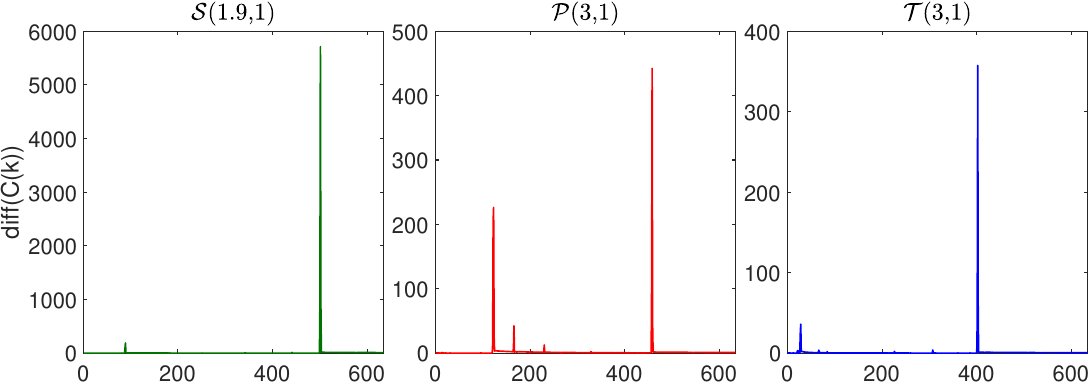} \\
        \includegraphics[width=0.95\textwidth]{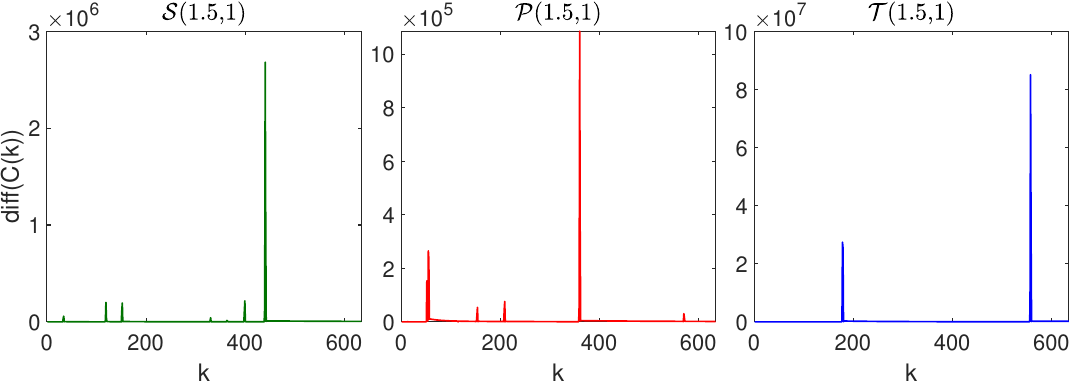}
        \caption{The increments of ECFM statistic calculated for the sub-signals demonstrated in Fig. \ref{fig_sub_signals}. The finite-variance cases (in TFD) are presented in the top row, while the infinite-variance cases (in TFD) are demonstrated in the middle and bottom row.}
    \label{incr_ecfm}
    \end{minipage}
    % \hfill
    \begin{minipage}[t]{0.45\textwidth}
        \centering
        \includegraphics[width=0.95\textwidth]{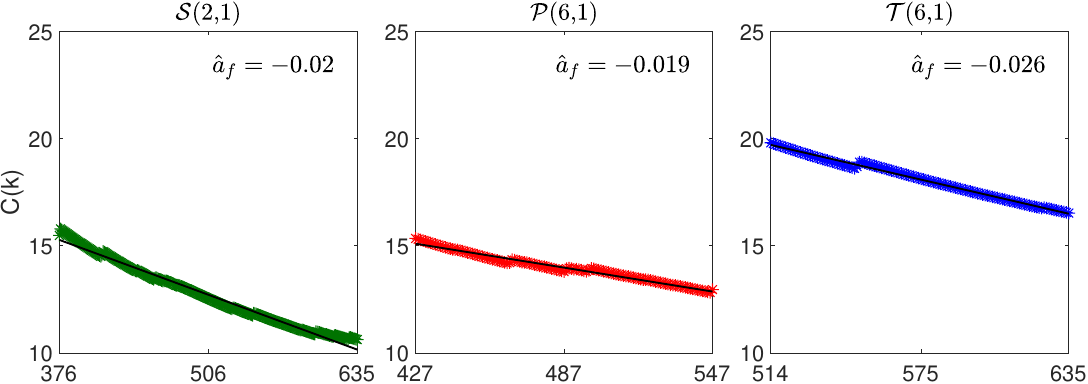} \\ 
        \includegraphics[width=0.95\textwidth]{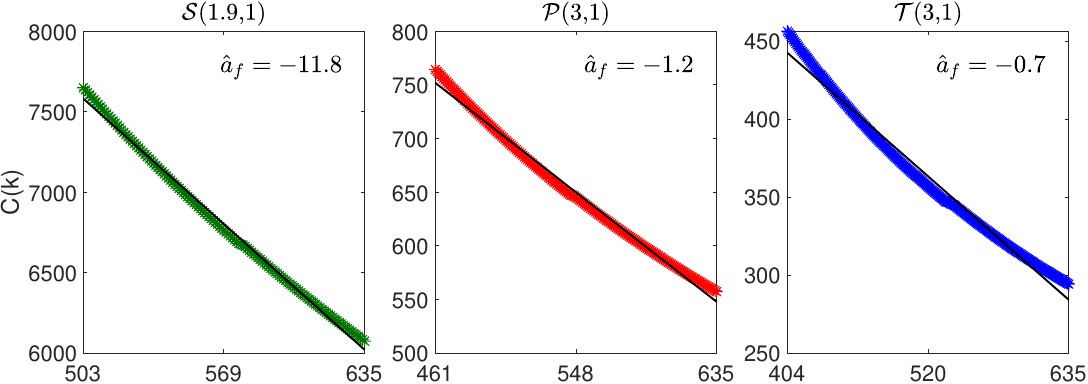} \\
        \includegraphics[width=0.95\textwidth]{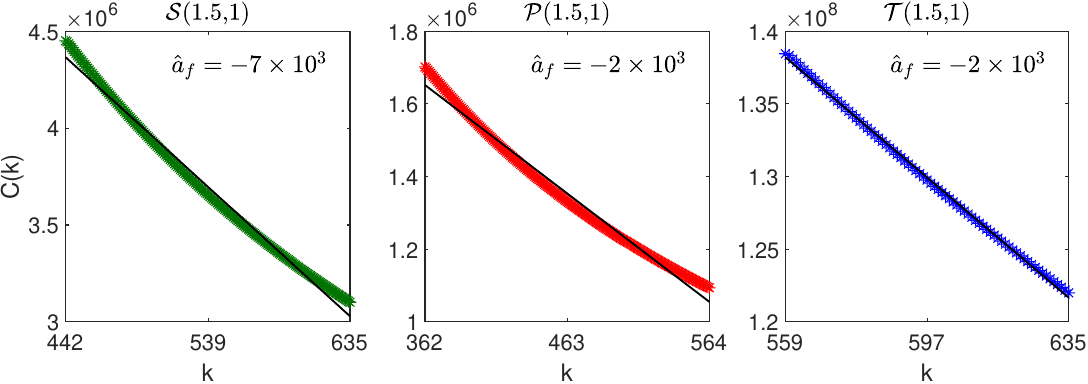}
        \caption{Selected segments of ECFM statistic calculated for the normalized sub-signals demonstrated in Fig. \ref{fig_sub_signals} and the fitted straight lines (black line) with the estimated slopes $a_f$. The finite-variance cases (in TFD) are presented in the top row, while the infinite-variance cases (in TFD) are demonstrated in the middle and bottom row.}
    \label{segment}
    \end{minipage}
\end{figure}
    
In Fig. \ref{slopy} we present the distribution of the slopes $a_f$ for simulated signals.  For each considered distribution and each set of parameters, we simulate signals of length $10000$. For each simulated signal, we apply the presented above procedure and calculate the median of the obtained $a_f$ values.  In Fig.  \ref{slopy} we present the distribution of the medians calculated for $1000$ simulated signals from the considered cases. In order to highlight the differences between considered distributions, the results are presented in the same scale on x-axis. In addition, in  Fig. \ref{slopy_a} we present the same results with the same scales on x-axis in the rows. One can clearly conclude about the significant differences between the values of $a_f$ estimated for distributions with finite- and infinite- variances.

\begin{figure}[H]
    \centering
    \begin{minipage}[t]{0.45\textwidth}
        \centering
   \includegraphics[width=.95\textwidth]{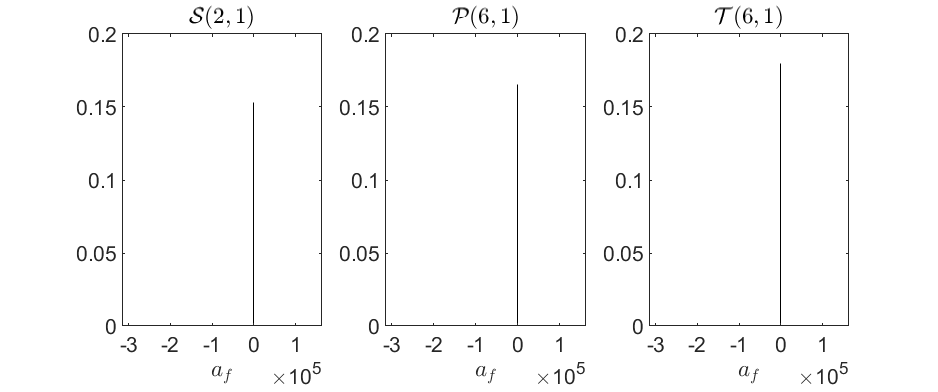}
    \\ \includegraphics[width=.95\textwidth]{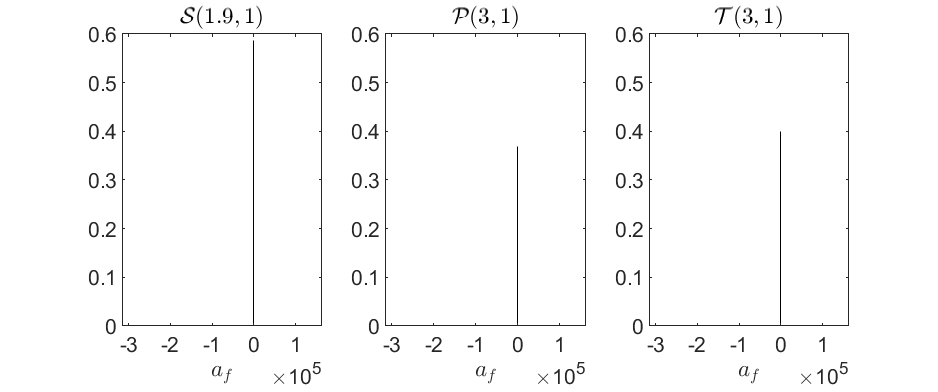}\\ \includegraphics[width=.95\textwidth]{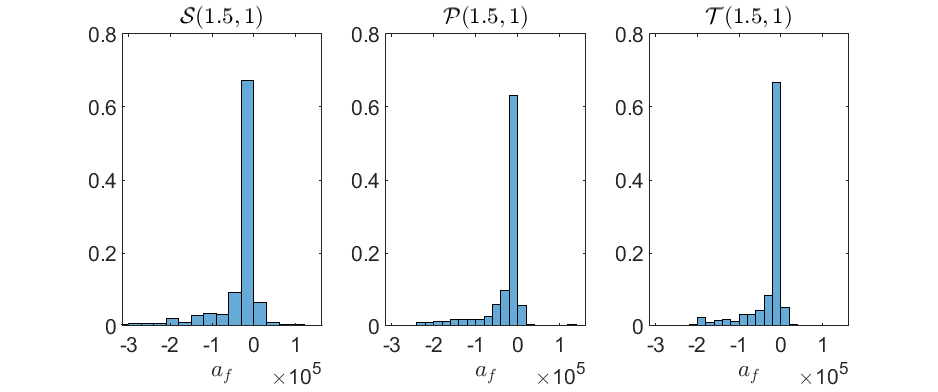}
        \caption{The distribution of the slopes $a_f$ for simulated signals from three considered distributions. Here we present the medians of the $a_f$ values obtained based on $1000$ simulated signals. To highlight the significant differences between the considered cases, the results are presented on the same scales on x-axis. }
    \label{slopy}
    \end{minipage}
    % \hfill
    \begin{minipage}[t]{0.45\textwidth}
        \centering
   \includegraphics[width=.95\textwidth]{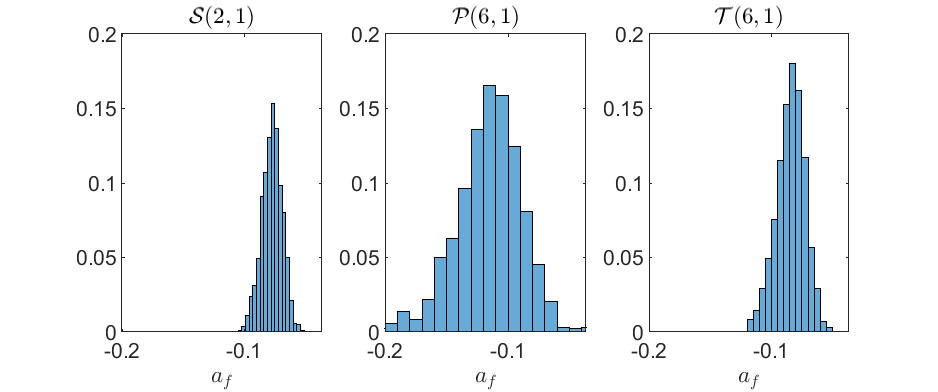}
    \\ \includegraphics[width=.95\textwidth]{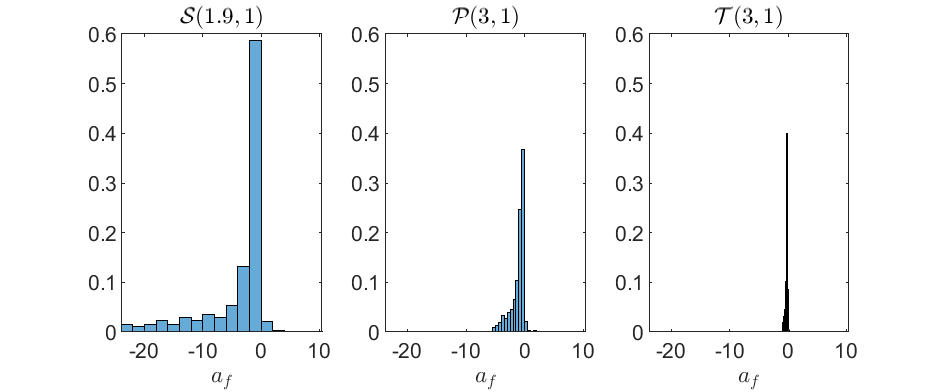}\\ \includegraphics[width=.95\textwidth]{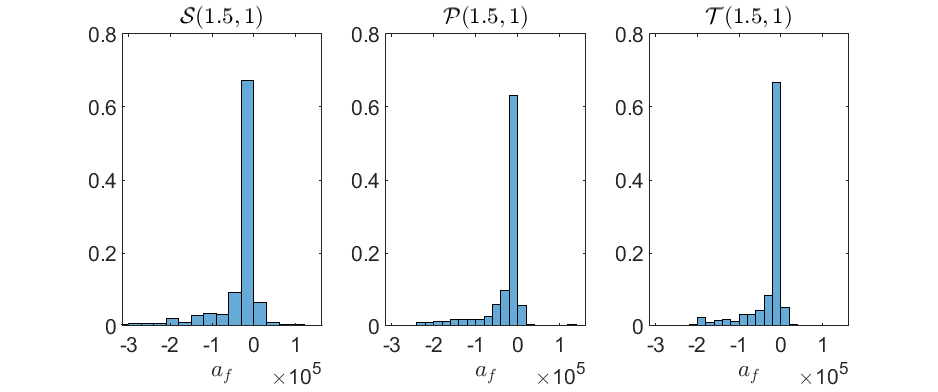}
        \caption{The distribution of the slopes $a_f$ for simulated signals from three considered distributions. Here we present the medians of the $a_f$ values obtained based on $1000$ simulated signals. The same scales on x-axis are applied for the cases presented in the same rows. }
    \label{slopy_a}
    \end{minipage}
\end{figure}

In order to underline the differences between the considered cases, in Fig. \ref{slopy_2} we present the comparison  of the medians of the estimated slopes $a_f$ presented in Fig. \ref{slopy} (left panel) and their IQRs, see right panel of Fig. \ref{slopy_2}. Because the  estimated slopes may be negative, here we present their absolute values to demonstrate the plots in log scales. In each case, the first bars correspond to the signals from finite-variance distribution in T-D, namely $\mathcal{S}(2,1)$, $\mathcal{P}(6,1)$ and $\mathcal{T}(6,1)$, presented in green, red and blue colors, respectively. The second bars correspond to the intermediate cases, namely $\mathcal{S}(1.9,1)$, $\mathcal{P}(3,1)$ and $\mathcal{T}(3,1)$ distributions.  The last bars correspond to $\mathcal{S}(1.5,1)$, $\mathcal{P}(1.5,1)$ and $\mathcal{T}(1.5,1)$ distributions. 

It is clearly seen that the medians of the estimated $a_f$ values are significantly smaller for the finite-variance distributed signals (first and second bars). For  $\mathcal{S}(2,1)$, $\mathcal{P}(6,1)$ and $\mathcal{T}(6,1)$ distributions, they are negative (in log scales). For the most extreme cases (last bars), the medians of the fitted slopes are significantly higher than for other cases. The differences between  the fitted slopes are also visible in the IQR statistic, considered as the dispersion measure. For the infinite-variance cases, the values of IQR are significantly higher than for finite-variance distributed signals. The medians and IQR of $a_f$ values clearly indicate the chaotic behavior of ECFM statistic for infinite-variance cases. 

\begin{figure}[H]
    \centering
   \includegraphics[width=.4\textwidth]{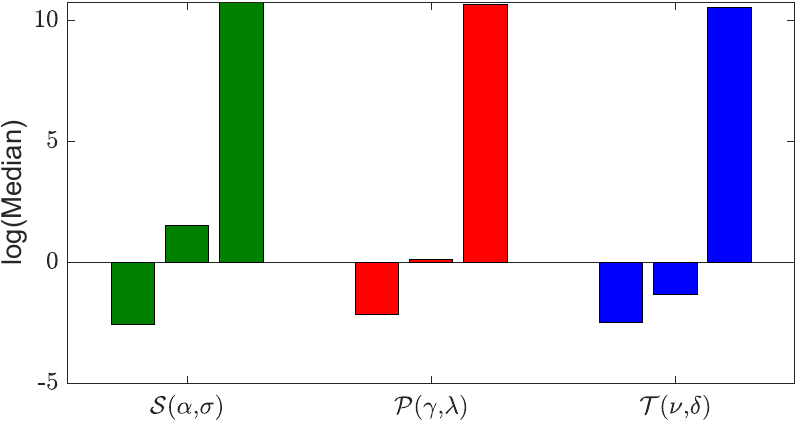} \includegraphics[width=.4\textwidth]{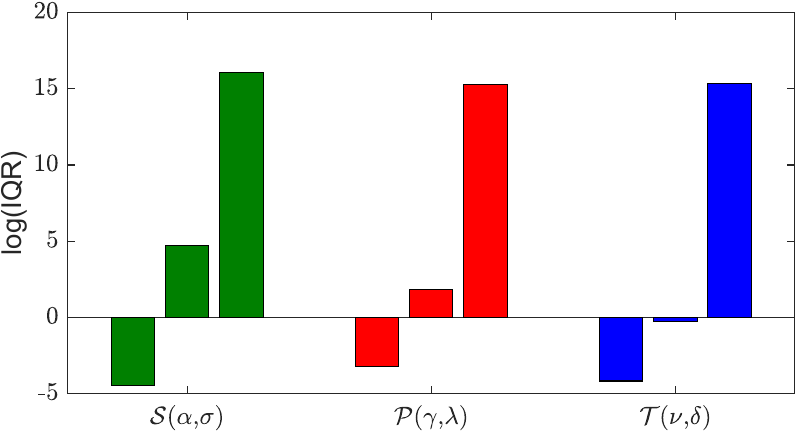}
    \caption{The comparison  of the medians of the estimated slopes $a_f$ presented in Fig. \ref{slopy} (left panel) and their interquartile ranges (IQR), see right  panel.  Here we present the absolute values and demonstrate the plots in log scales. In each case, the first bars correspond to the signals from finite-variance distributions, the second bars correspond to the intermediate cases, and the last bars correspond to infinite-variance distributions in TFD. The green color corresponds to $\alpha-$stable distribution, the red one to the symmetric Pareto distribution, the blue one - to t location-scale distribution.}\label{slopy_2}
\end{figure}

\section{Real signals analysis}\label{real}
In this section, we present the analysis for four real signals with potential different distributions of the background noise, see Fig. \ref{real1}. {The signals correspond to the healthy machines, thus we do not expect here any signal of interest. Moreover, the high-pass filtering was applied for all considered signals to remove the possible deterministic components.}

To illustrate the problem, we propose to investigate four exemplary signals denoted  here sig. 1, sig. 2, sig. 3, and sig. 4. These signals come from various machines, namely, vibration signal from rolling element bearings from pulley used in belt conveyor system (sig. 1), acoustic signal (sound) measured close to idler installed in belt conveyor (sig. 2), vibration signals from hammer crusher used in copper ore processing plant for hard rock material fragmentation (sig. 3 and sig. 4).

The first considered signal (denoted further as sig. 1) describes vibration from healthy bearings, there is an amplitude modulation with cycle c.a. 1$Hz$ that is related to some minor shaft problems. No impulsive behaviour is present in the time series.

The second signal (sig. 2) presents noise signal from healthy bearings in idler. Minor impulses are visible in the time domain. However, in that example, signature of the healthy element is contaminated by several high amplitude impulses from moving clamp connecting two parts of the belt (clearly seen on spectrogram presented in Fig. \ref{real_spectr}b).

Sig. 3 and sig. 4 describe vibration of crushers. The difference between signals is related to operational conditions (load, i.e., material stream coming to the crusher). As granulation of copper ore fed to machine may be very different (from sand like material to pieces of rocks with a dozen of kilograms) signals may contain nearly Gaussian noise or due to shocks - a strongly impulsive components. 

It was already mentioned, that first look at the real signal may be a bit confusing. It may pretend to be Gaussian distributed, however wideband impulsive behavior may be hidden in the measured data. Thus, to investigate true properties of real signals we use time-frequency representation (spectrogram), see Fig. \ref{real_spectr}.  To calculate spectrograms we have applied the 'spectrogram' Matlab procedure with the following parameters: for sig. 1 and sig. 2 window $kaiser(500,5)$, overlap size $474$ and $512$ points to calculate fast Fourier transform, for sig. 3 and sig 4 window $kaiser(2000,5)$, overlap size $1896$ and $2048$ points to calculate fast Fourier transform. 

As it is clearly seen, time-frequency representations of the analyzed  signals are different, as machines are different. However, there are some important features. Some frequency bands with high energy (expressed as red color) are usually present at low frequencies. For high frequencies, the energy is smaller (sig. 1), however some resonances may appear (sig. 3, sig. 4). Non-Gaussian heavy-tailed behavior is related to vertical lines  that means wideband (impulsive) disturbance existing at some time instance. This is clearly seen for sig. 2.  
 
 \begin{figure}[H]
  \begin{subfigure}[t]{.2125\textwidth}
    \centering
    \includegraphics[width=\linewidth]{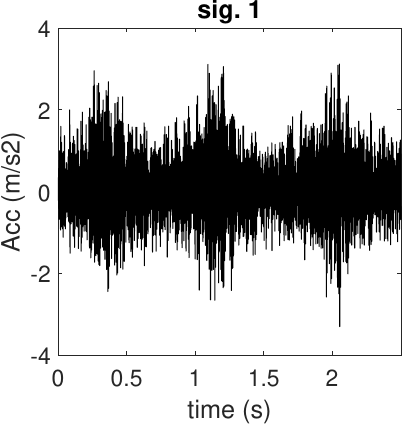}
   % \caption{lozysko.mat}
  \end{subfigure}
  %\hspace{1em}
  \centering
  \begin{subfigure}[t]{.2\textwidth}
    \centering
    \includegraphics[width=\linewidth]{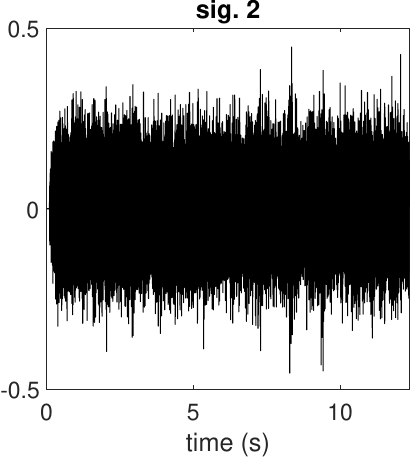}
  %  \caption{dobry.mat}
  \end{subfigure}
  \medskip
  \begin{subfigure}[t]{.2\textwidth}
    \centering
    \includegraphics[width=\linewidth]{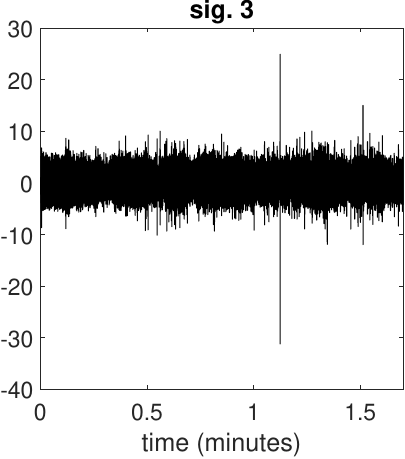}
%    \caption{sig14.mat - kruszarka}
  \end{subfigure}
  %\hspace{1em}
  \centering
  \begin{subfigure}[t]{.2\textwidth}
    \centering
    \includegraphics[width=\linewidth]{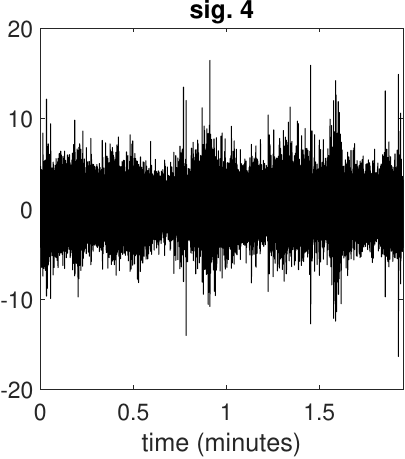}
 %   \caption{sig15.mat - kruszarka}
  \end{subfigure}
  \caption{The analysed signals in TD.}\label{real1}
\end{figure}

\begin{figure}[H]
  \begin{subfigure}[t]{.2225\textwidth}
    \centering
    \includegraphics[width=\linewidth]{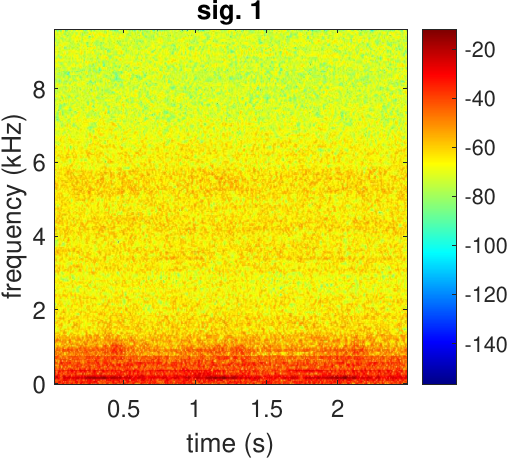}
    %\caption{Spectrogram lozysko.mat}
  \end{subfigure}
  %\hspace{1em}
  \centering
  \begin{subfigure}[t]{.2125\textwidth}
    \centering
    \includegraphics[width=\linewidth]{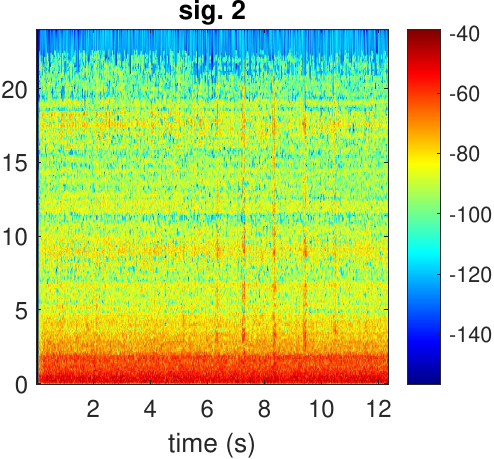}
  %  \caption{Spectrogram dobry.mat}
  \end{subfigure}
  \medskip
  \begin{subfigure}[t]{.2125\textwidth}
    \centering
    \includegraphics[width=\linewidth]{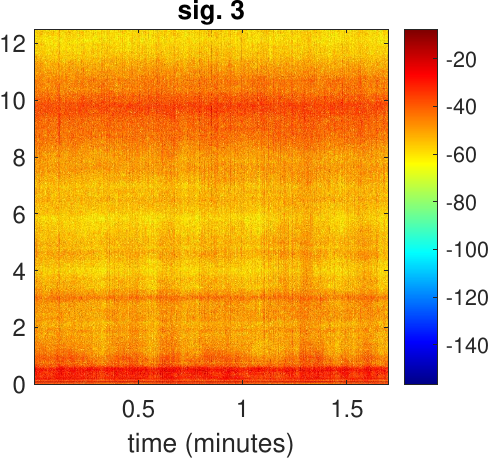}
   % \caption{Spectrogram sig14.mat - kruszarka}
  \end{subfigure}
  %\hspace{1em}
  \centering
  \begin{subfigure}[t]{.225\textwidth}
    \centering
    \includegraphics[width=\linewidth]{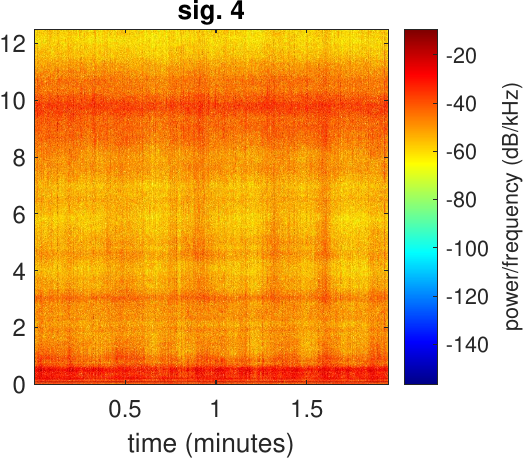}
   % \caption{Spectrogram sig15.mat - kruszarka}
  \end{subfigure}
  \caption{Spectrograms of the signals demonstrated in Fig. \ref{real1}.}\label{real_spectr}
\end{figure}

In order to identify the possible infinite-variance distribution of the real signals, we apply the procedure described in the previous section. More precisely, for each signal, we analyze the ECFM statistics for sub-signals taken from the spectrogram.  Then, we  parametrize the possible chaotic behavior of ECFM statistics by analyzing the estimated slopes $a_f$ from the linear regression applied for the last long segment of ECFM. In Fig. \ref{real2} we demonstrate the distribution of the obtained $a_f$ values along frequencies. To the analysis, we take frequencies from $4.5kHz$ to $9kHz$ for sig. 1, $12.5kHz$ to $22.5kHz$ for sig. 2, $6kHz$ to $12.5kHz$ for sig. 3 and sig. 4. {It is worthy to note, that due to significant differences between sub-signals associated with different frequency bins, we are not allowed to pick a single vector to test the presence of non-Gaussianity. Thus, we are doing this for a wide range of frequencies, and we have found similarities between these sub-signals. In this way, we minimize the probability of  misleading decisions.}
\begin{figure}[H]
  \begin{subfigure}[t]{.2\textwidth}
    \centering
    \includegraphics[width=\linewidth]{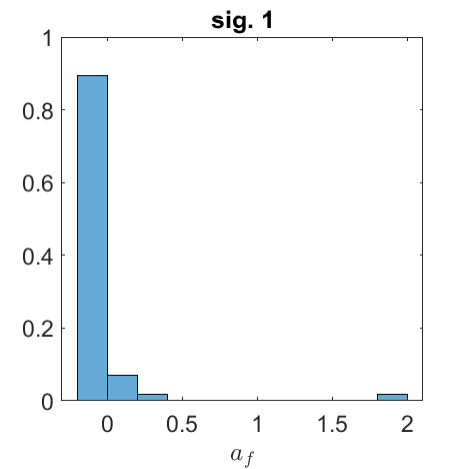}
   % \caption{lozysko.mat}
  \end{subfigure}
  \hspace{1em}
  \centering
  \begin{subfigure}[t]{.2\textwidth}
    \centering
    \includegraphics[width=\linewidth]{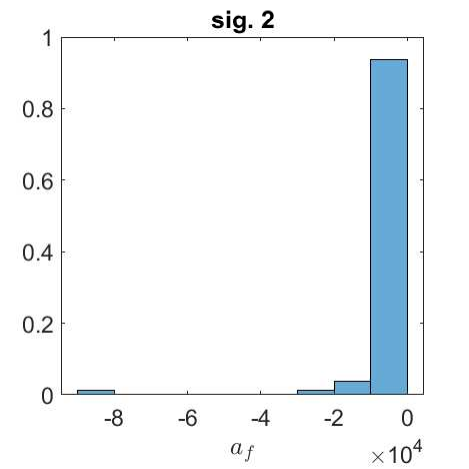}
  %  \caption{dobry.mat}
  \end{subfigure}
  \medskip
  \begin{subfigure}[t]{.2\textwidth}
    \centering
    \includegraphics[width=\linewidth]{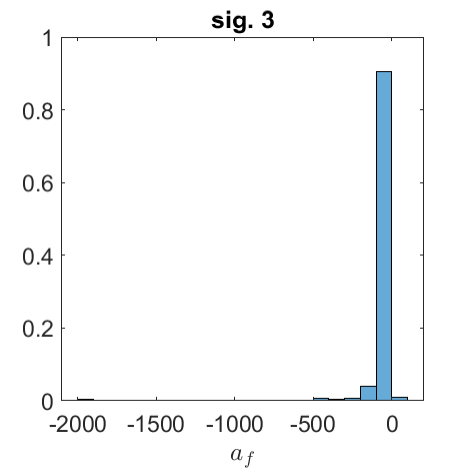}
  %  \caption{sig14.mat - kruszarka}
  \end{subfigure}
  \hspace{1em}
  \centering
  \begin{subfigure}[t]{.2\textwidth}
    \centering
    \includegraphics[width=\linewidth]{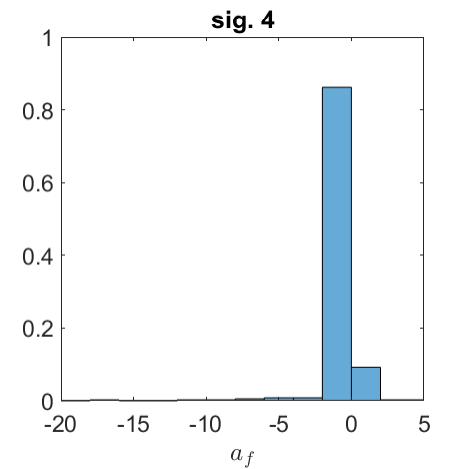}
  %  \caption{sig15.mat - kruszarka}
  \end{subfigure}
  \caption{The distribution of the $a_f$ values for four analyzed signals along the frequencies.}\label{real2}
\end{figure}
Analyzing the results presented in Fig. \ref{real2}, we can conclude that for sig. 2  and sig. 3 we expect the chaotic behavior of ECFM statistic applied for the signals in TFD. Thus, we can assume the infinite-variance distributions of the background noise, similar as for the signals presented in the bottom panel in Fig. \ref{fig1}. Obviously, significantly smaller values of $a_f$ are obtained in sig. 1, when we expect the finite-variance distribution of the signal in TFD. This signal may correspond to the cases presented in the top panel of Fig. \ref{fig1}. For sig. 4 the values of the estimated slopes are higher than for sig. 1 but significantly smaller than for sig. 2 and sig. 3. Thus, one may conclude that this signal corresponds to the intermediate case (corresponding to cases presented in the middle panel of Fig. \ref{fig1}) or to the finite-variance distribution but with a heavier tail than in sig. 1.  To confirm our preliminary assumption about the classes of distributions corresponding to the analyzed signals in TFD, in Fig. \ref{real3} we demonstrate medians (left panel) and IQRs (right panel) for the estimated values of $a_f$ calculated from the estimated slopes for all frequencies. Similar as for simulated signals, here we demonstrate the logarithm of the descriptive statistics applied for the absolute values of $a_f$.

\begin{figure}[H]
    \centering
    \includegraphics[width=.4\textwidth]{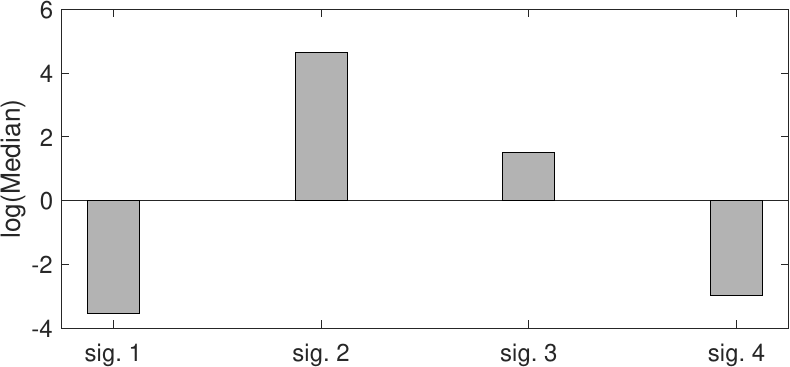}
    \includegraphics[width=.4\textwidth]{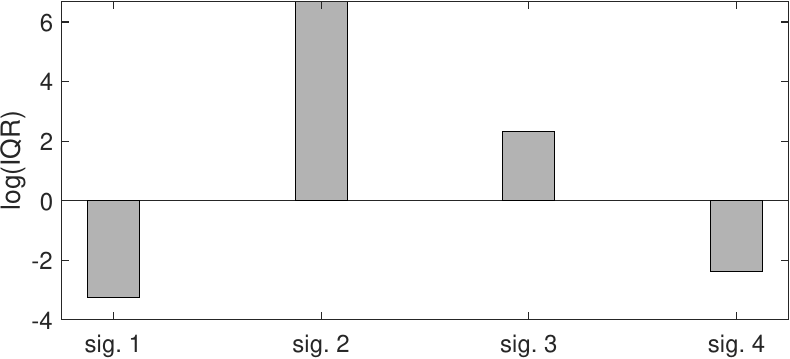}
    \caption{Results for real signals. Medians (left panel) and IQRs (right panel) for the estimated values of $a_f$ calculated from the estimated slopes for all frequencies. Here we demonstrate the log of the descriptive statistics applied for the absolute values of $a_f$. }\label{real3}

\end{figure}

Our preliminary assumptions about the identification of the distribution class is confirmed by Fig. \ref{real3}. Comparing the results presented for simulated signals, see Fig. \ref{slopy_2}, we can classify sig. 1 as the finite-variance distributed. Sig. 2 and sig. 3 are classified as infinite-variance distributed signals. However, for sig. 3 the distribution tail is lighter than for sig. 2. Sig. 4 corresponds finite-variance case. The last step of our analysis is the comparison of the empirical tails of the selected sub-signals taken from the spectrograms for real signals with the tails of the fitted generalized $\chi^2$ distribution (see Eq. (\ref{chi2pdf})). The parameters of the generalized $\chi^2$ distribution taken to the comparison are estimated from the analyzed sub-signals. The results for selected sub-signals from the corresponding spectrograms are presented in Fig. \ref{real4}. The green line corresponds to the generalized $\chi^2$ distribution, while the black stars present the empirical CDFs. The results are presented in log-log scales. For the analysis, we selected the sub-signals arbitrary. However, for other sub-signals we obtain similar results. To confirm that all sub-signals taken from the spectrograms correspond to the same classes in  Fig. \ref{appb} (see \ref{ap_a}) we demonstrate the  empirical tails for all sub-signals.

\begin{figure}[H]
  \begin{subfigure}[t]{.2\textwidth}
    \centering
    \includegraphics[width=\linewidth]{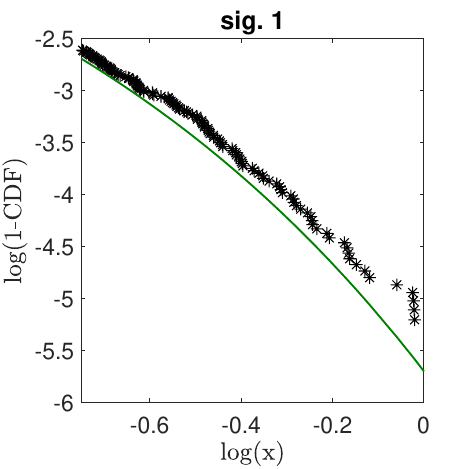}
    
  \end{subfigure}
  \hspace{1em}
  \centering
  \begin{subfigure}[t]{.2\textwidth}
    \centering
    \includegraphics[width=\linewidth]{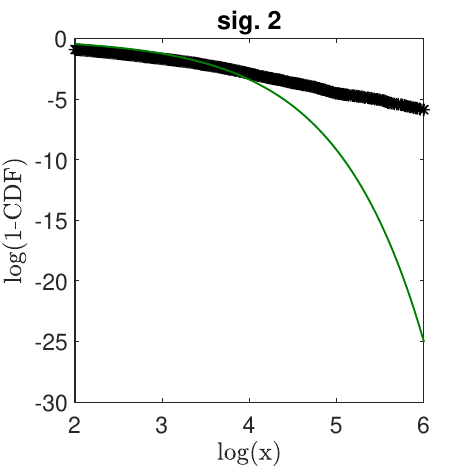}
   
  \end{subfigure}
  \medskip
  \begin{subfigure}[t]{.2\textwidth}
    \centering
    \includegraphics[width=\linewidth]{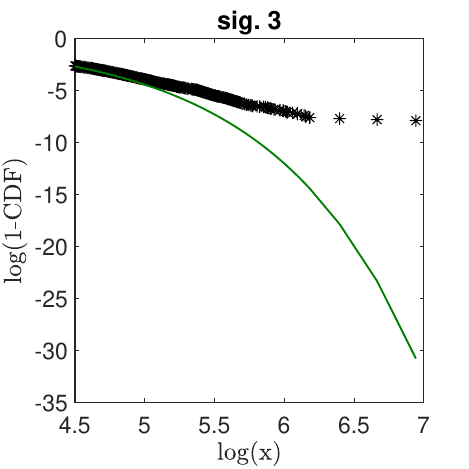}
    
  \end{subfigure}
  \hspace{1em}
  \centering
  \begin{subfigure}[t]{.2\textwidth}
    \centering
    \includegraphics[width=\linewidth]{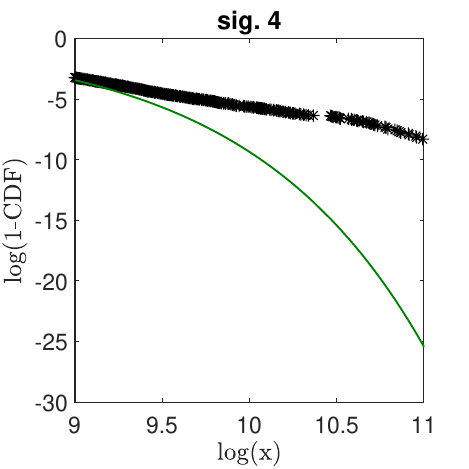}
   
  \end{subfigure}
  \caption{The comparison of the empirical tails of the selected sub-signals taken from the spectrograms for real signals with the tails of the fitted generalized $\chi^2$ distribution with estimated parameters. The green line corresponds to the generalized $\chi^2$ distribution, while the black stars present the empirical CDFs.}\label{real4}
\end{figure}

Analyzing Fig. \ref{real4}, we can conclude that the selected sub-signal corresponding to sig. 1 may be considered as generalized $\chi^2$ distributed, while the sub-signals corresponding to sig. 2, sig. 3 and sig. 4 have heavier tails. We note, sig. 4 however is finite-variance distributed in TFD. The assumption of the generalized $\chi^2$ distribution for sig. 1  is confirmed by KS test. The median of $p_{value}$ of KS test for generalized $\chi^2$ distribution applied to all sub-signals from the  spectrograms for sig. 1 is equal to $0.27$.  The $p_{value}$s for all sub-signals are higher than $10\%$. We remind, the $p_{value}$ higher than the confidence level here $5\%$ indicates the hypothesis of tested distribution can not be rejected. For other signals, the $p_{value}$s of KS test are significantly smaller than $5\%$. The boxplots of  $p_{value}$s of the KS test for all considered signals are presented in  Fig. \ref{box} in \ref{ap_a}. The practical aspect of the obtained results is simple, namely for sig. 1 and sig. 4 the  approaches based on the classical auto-dependence measures can be applied for local damage detection, while for sig. 2 and sig. 3 they may be not useful. For this case, more appropriate tools, dedicated to infinite-variance distributed signals, need to be applied for the analysis of the signals in TFD.

\section{Conclusions}

{In many cases, researchers working on local damage detection pay attention on properties of the signal of interest, but not of the background noise}. In this paper, we have built another perspective. We note that almost all measured signals are associated with some noise. Thus, before applying methods for damage detection that are based on two of the most intuitive features of the SOI, namely impulsiveness or cyclic/periodic nature, one has to check if the use of a given  algorithm is allowed by the properties of the background noise.  In this paper,  we discuss  the probabilistic properties of the background noise and indicate that they are extremely important in the context of applying classical methods for damage detection. The noise may have a Gaussian or non-Gaussian distribution. Even if we identify the non-Gaussian distribution of the background noise, this information may not be sufficient for selection of proper tools for signal analysis. 
    In this paper, we categorize the types of noise depending on the existence of the variance of its distribution. The selection of the variance as the criterion is related to the fact that most of the classical techniques used for local damage detection actually require it to be finite. 
    {The problem and the proposed methodology are demonstrated based on three popular non-Gaussian distributions that are used as  models of impulsive noise. We discuss the problem of non-Gaussian heavy-tailed distribution of the background noise in the time and the time-frequency domains,  as many techniques are applied in these domains. We have demonstrated that the non-Gaussian character of the noise in the time domain is transferred to the time-frequency domain, however the level of non-Gaussianity increases through applying squared STFT, i.e. the spectrogram. In consequence, the finite-variance distributed signal may become infinite-variance distributed after the transformation to the other domain.  This observation sheds new light on the application of classical methods to signal analysis in the time and time-frequency domains.  
    
    As a main solution, we have proposed an adaptation of visual test based on the ECFM for variance presence testing. The methodology is based on the time-frequency representation of a given signal and takes under consideration the specific behavior of used statistics for infinite-variance distributed data. The proposed methodology  is intuitive, simple to interpret and efficient, i.e. it provides clear information if classical methodology for SOI detection can be applied to the underlying signal. 
    
According to our previous research, even some classical techniques may be useful for damage detection in the presence of non-Gaussian noise. However, the classical methods fail if the non-Gaussianity level is significant, \cite{Wodecki2021}. In this paper, we continue this research and demonstrate how to identify the extreme cases when the classical approaches may not give reliable results. 
    We believe, that the proposed approach could be very helpful to all researchers working with noisy signals when the noise is non-Gaussian and heavy-tailed distributed.}

\section*{Acknowledgements}
The work of KS, RZ and AW is supported by the National Center of Science under Sheng2 project No. UMO-2021/40/Q/ST8/00024 "NonGauMech - New methods of processing non-stationary signals (identification, segmentation, extraction, modeling) with non-Gaussian characteristics for the purpose of monitoring complex mechanical structures". 
\bibliography{mojabib}
\appendix
\section{Additional figures}\label{ap_a}

\begin{figure}[H]
  \begin{subfigure}[t]{.2\textwidth}
    \centering
    \includegraphics[width=\linewidth]{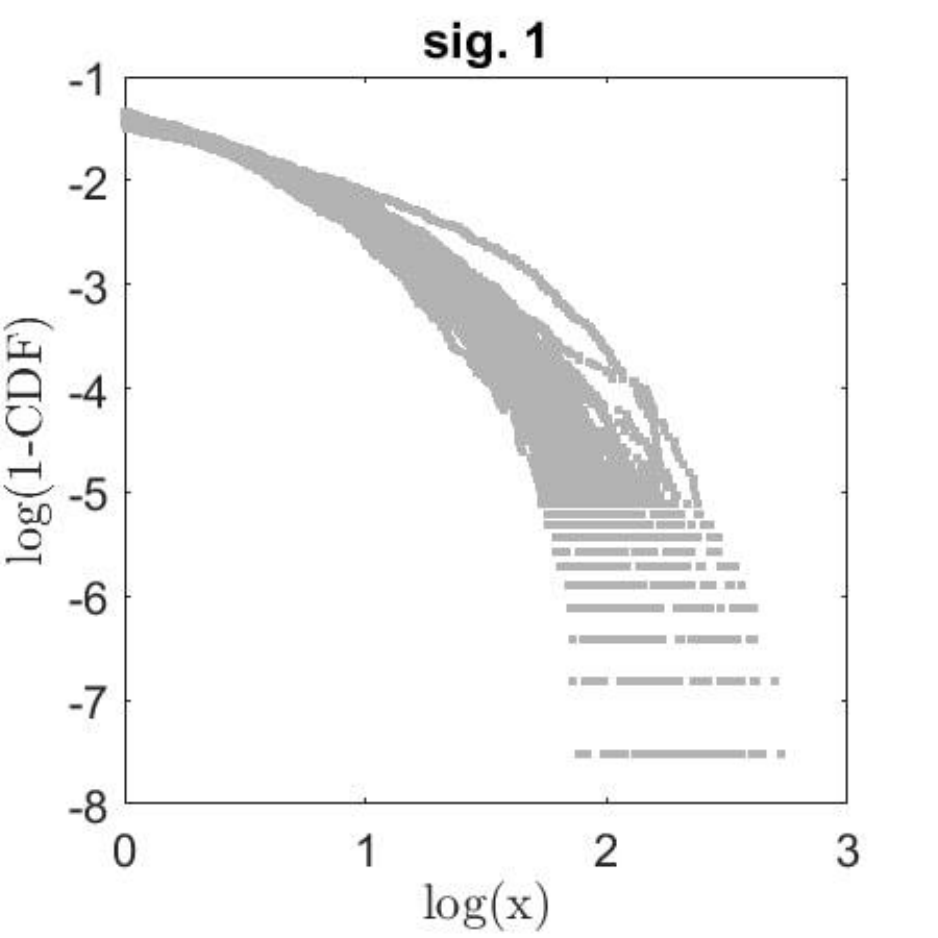}
  %  \caption{lozysko.mat}
  \end{subfigure}
  \hspace{1em}
  \centering
  \begin{subfigure}[t]{.2\textwidth}
    \centering
    \includegraphics[width=\linewidth]{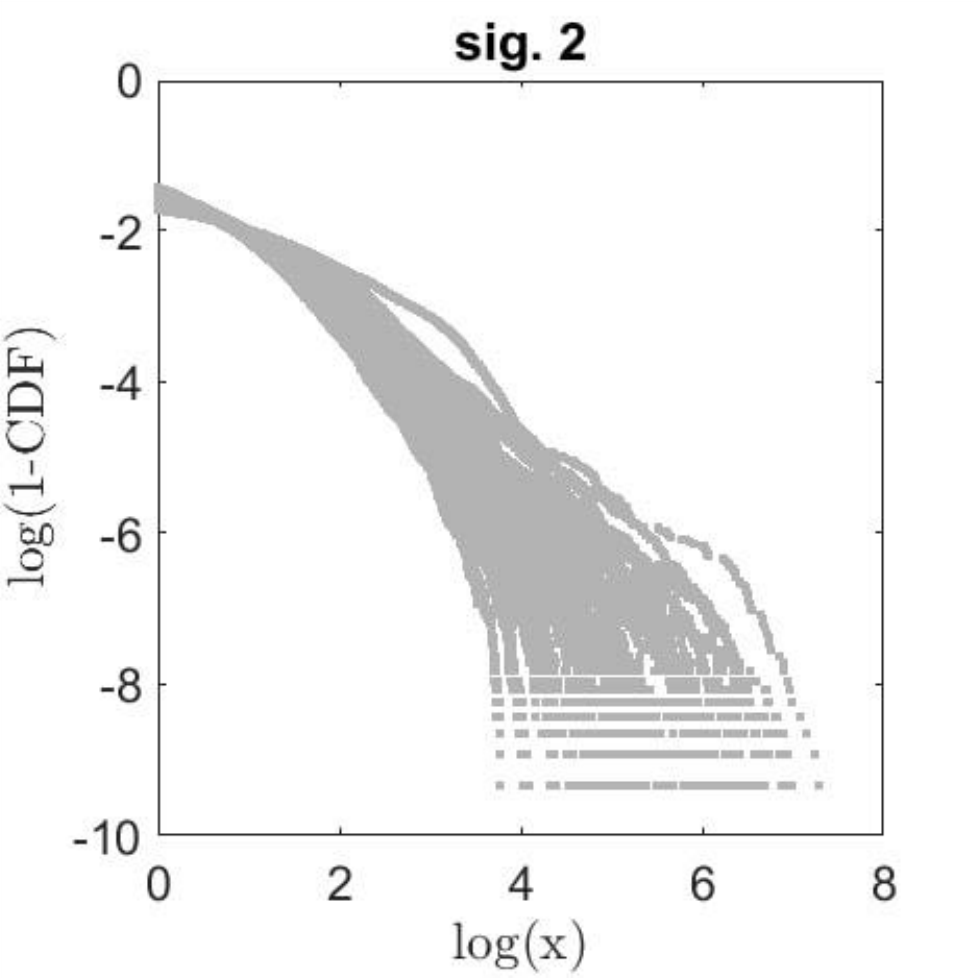}
   % \caption{dobry.mat}
  \end{subfigure}
  \medskip
  \begin{subfigure}[t]{.2\textwidth}
    \centering
    \includegraphics[width=\linewidth]{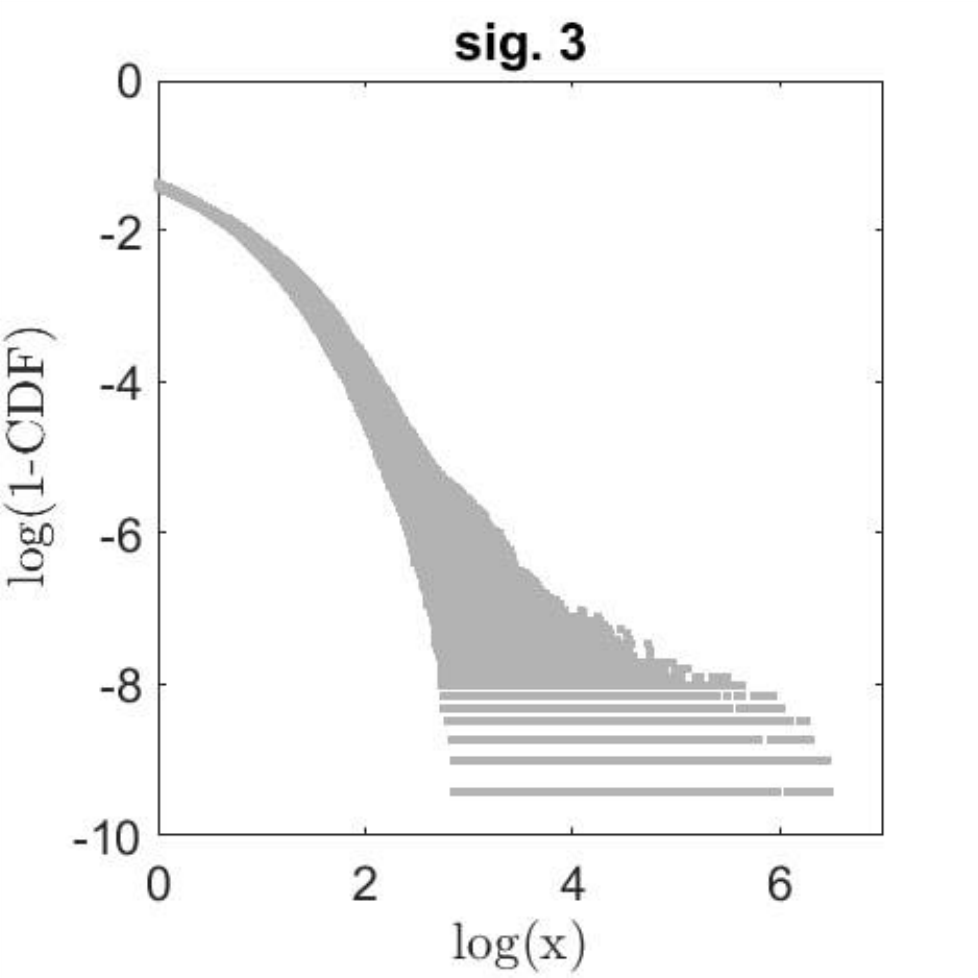}
    %\caption{sig14.mat - kruszarka}
  \end{subfigure}
  \hspace{1em}
  \centering
  \begin{subfigure}[t]{.2\textwidth}
    \centering
    \includegraphics[width=\linewidth]{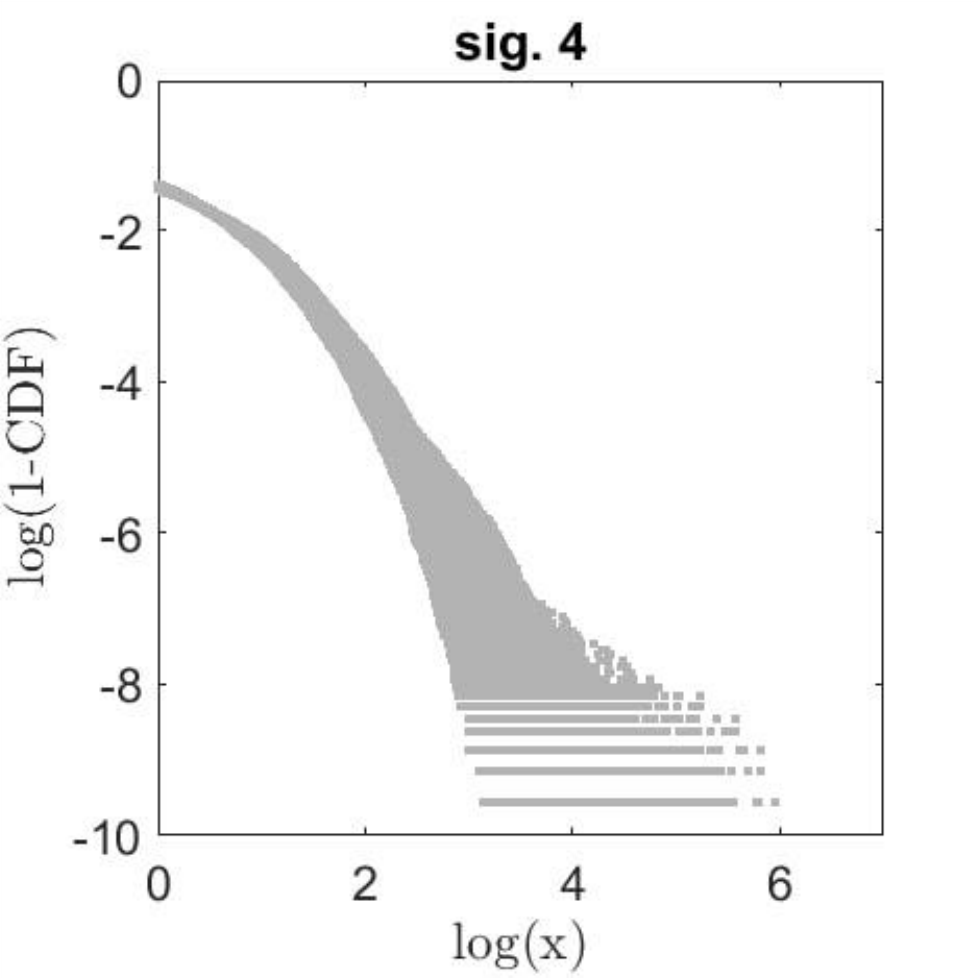}
    %\caption{sig15.mat - kruszarka}
  \end{subfigure}
  \caption{Empirical tails of the sub-signals taken from the spectrograms of the real signals. }\label{appb}
\end{figure}

\begin{figure}[H]
  \begin{subfigure}[t]{.2\textwidth}
    \centering
    \includegraphics[width=\linewidth]{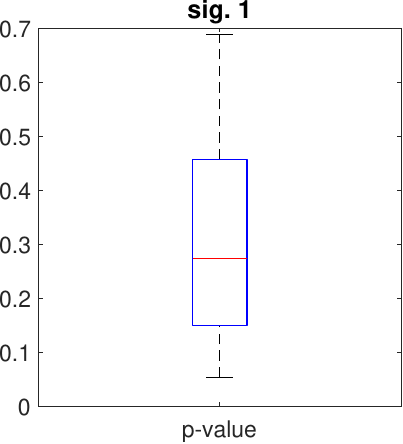}
    %\caption{lozysko.mat}
  \end{subfigure}
   \centering
  \begin{subfigure}[t]{.2125\textwidth}
    \centering
    \includegraphics[width=\linewidth]{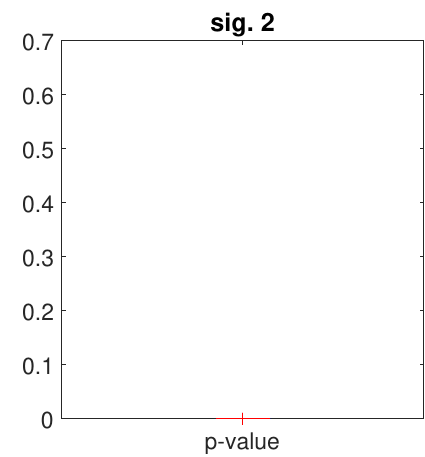}
    %\caption{niecykliczny noise.mat}
  \end{subfigure}
 \begin{subfigure}[t]{.2125\textwidth}
    \centering
    \includegraphics[width=\linewidth]{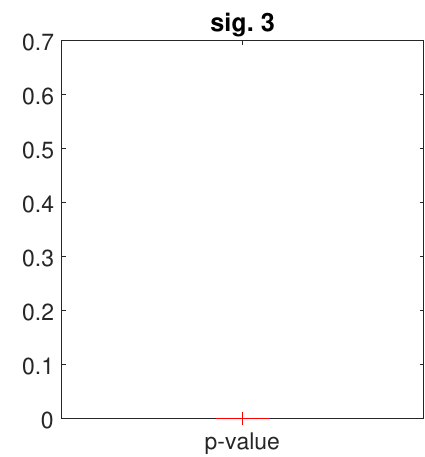}
    %\caption{sig14.mat - kruszarka}
  \end{subfigure}
  %\hspace{1em}
  \centering
  \begin{subfigure}[t]{.215\textwidth}
    \centering
    \includegraphics[width=\linewidth]{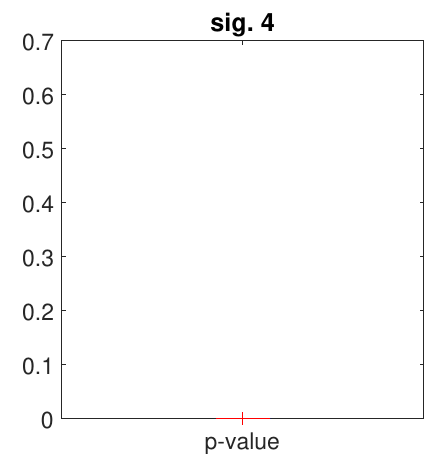}
    %\caption{sig15.mat - kruszarka}
  \end{subfigure}
  \caption{Boxplots of $p_{values}$s of KS test for $\chi^2$ distribution applied for the sub-signals from the spectrograms of analyzed real signals.}\label{box}
\end{figure}

\begin{algorithm}
\caption{Algorithm for identification of infinite variance of the signal}\label{alg:cap}
\begin{algorithmic}
%\Require $n \geq 0$
%\Ensure $y = x^n$
\State Take a signal in TD
\State Calculate $S(\cdot,\cdot)$
\State Normalize spectrogram for each $f\in \mathcal{F}$: $S_{norm}(\cdot,f) \gets (S(\cdot,f) - median(S(\cdot,f)))/\text{CondStd}(S(\cdot,f))$ 
\For{$f\in \mathcal{F}$}
\State Calculate ECFM for each $f\in \mathcal{F}$ for $S_{norm}(\cdot,f)$
\State Calculate increments of ECFM for $S_{norm}(\cdot,f)$ using diff()
\State Identify segments of ECFM based on values of increments.
\State Choose last segment longer than $10\%$ of initial data.
\State Calculate $\hat{a}_f$.
\EndFor
\State Analyze IQR and Median of calculated $\hat{a}_f$ for all $f\in \mathcal{F}$.
\label{algorithm}
\end{algorithmic}
\end{algorithm}

\end{document}